\documentclass[aps,onecolumn,groupedaddress,nofootinbib]{revtex4}
\usepackage{graphicx,color}
\usepackage{amsmath}
\usepackage{multirow}
\usepackage[caption=false]{subfig}
\usepackage{csquotes}

\begin{document}

\title{Analytical expressions for thermophysical properties of solid and liquid tungsten relevant for fusion applications}
\author{P. Tolias and the EUROfusion MST1 Team\footnotemark\footnotetext{\scriptsize See author list of \enquote{H. Meyer \emph{et al.} 2017 Nucl. Fusion 57 102014}}}
\date{\today}
\affiliation{Space and Plasma Physics, Royal Institute of Technology, Stockholm, Sweden}
\begin{abstract}
\noindent The status of the literature is reviewed for several thermophysical properties of pure solid and liquid tungsten which constitute input for the modelling of intense plasma-surface interaction phenomena that are important for fusion applications. Reliable experimental data are analyzed for the latent heat of fusion, the electrical resistivity, the specific isobaric heat capacity, the thermal conductivity and the mass density from the room temperature up to the boiling point of tungsten as well as for the surface tension and the dynamic viscosity across the liquid state. Analytical expressions of high accuracy are recommended for these thermophysical properties that involved a minimum degree of extrapolations. In particular, extrapolations were only required for the surface tension and viscosity.
\end{abstract}
\maketitle

\section{Introduction}

\noindent The survivability of the divertor during prolonged repetitive exposures to harsh edge plasma conditions as well as its longevity deep into the nuclear phase are essential for the success of the ITER project and impose stringent requirements on material selection\,\cite{introduct00}. After the decision that ITER will begin operations with a full tungsten divertor\,\cite{introduct01}, R\&D activities worldwide have focused on assessing various sources of mechanical and structural degradation of tungsten plasma-facing components in the hostile fusion reactor environment\,\cite{introduct02,extraintro1,extraintro2}; neutron irradiation effects on mechanical properties, helium irradiation and hydrogen retention effects on the microstructure, thermal shock resistance and thermal fatigue resistance. The dependence of key mechanical properties (ductile to brittle transition temperature, yield strength, fracture toughness) on the fabrication history, the alloying or impurity elements, the metallurgical process and the grain structure is indicative of the complex nature of such investigations\,\cite{introduct03,introduct04}. As a consequence, the ITER Materials Properties Handbook puts strong emphasis to documenting the mechanical properties of tungsten\,\cite{introduct05}.

Another phenomenon that is crucial for the lifetime of the tungsten divertor is melt layer motion during off-normal or transient events, namely unmitigated edge localized modes, vertical displacement events and major disruptions\,\cite{introduct01}. Melt layer motion leads to strong modifications of the local surface topology and thus to degradation of power-handling capabilities but can also lead to plasma contamination by high-Z droplets in the case of splashing\,\cite{introduct06,introduct07}. The numerical modelling of melt motion is based on coupling the Navier-Stokes equations for the liquid metal with the heat conduction equation as well as the current continuity equation and supplementing the system with appropriate boundary conditions dictated by the incident plasma\,\cite{introduct08,introduct09,introduct10}. These are the fundamental equations solved in codes such as MEMOS\,\cite{introduct08,introduct09}, where the temperature dependence of the viscosity, surface tension and other thermophysical properties of liquid tungsten constitute a necessary input. In addition, since re-solidification determines the onset but also the arrest of macroscopic motion, the thermophysical properties of solid tungsten at elevated temperatures and their behavior at the solid-liquid phase transition are also necessary input. Unfortunately, the ITER Materials Properties Handbook does not provide any information on the thermophysical properties of liquid tungsten and its recommended description of some thermophysical properties of solid tungsten at elevated temperatures is not based on state-of-the-art experimental data\,\cite{introduct05}. It should also be mentioned that these properties are also essential input for the modelling of tungsten dust transport with codes such as DUSTT\,\cite{introduct11} and MIGRAINe\,\cite{introduct12} (since tungsten dust should promptly melt in ITER-like edge plasmas and thermionic emission at the liquid phase plays a dominant role in its energy budget) and for the modelling of the interaction of transient plasmas with adhered tungsten dust\,\cite{introduct13} (since wetting is determined by the competition between the spreading and re-solidification rates).

This work is focused on reviewing state-of-the-art measurements of thermophysical properties of pure tungsten from the room temperature up to the boiling point. Complications arising in fusion devices due to strong magnetic fields, intense plasma fluxes, impurity alloying and neutron irradiation are also discussed. The thermophysical properties of interest are the latent heat of fusion, the electrical resistivity, the specific isobaric heat capacity, the thermal conductivity and the mass density (solid and liquid phase) as well as the surface tension and the dynamic viscosity (liquid phase). The objective is to identify and critically evaluate reliable experimental datasets in order to propose accurate analytical expressions for the temperature dependence of these quantities that will standardize their description in the multiple heating, melt layer motion and dust transport codes developed by the fusion community. It has been possible to provide accurate analytical expressions for most properties based solely on experimental data and without the need for any extrapolations. The only exceptions are the surface tension and viscosity of liquid tungsten, where wide extrapolations had to be carried out beyond the melting point, since the only experimental sources on the temperature dependence referred to the under-cooled phase. These extrapolations are based on established empirical expressions that are accurate for non-refractory liquid metals and were cross-checked with rigorous constraints imposed by statistical mechanics. Considering that temperature gradients of the surface tension can drive thermo-capillary flows and that viscosity is responsible for melt motion damping, measurements need to be carried out in the unexplored temperature range, e.g. with levitating drop methods on ground-based laboratories\,\cite{introduct14} or in microgravity\,\cite{introduct15}.

\section{Thermophysical properties of tungsten}

\subsection{The latent heat of fusion}\label{latent}

\noindent The difference between the specific enthalpy of the liquid and solid state at the melting phase transition yields the latent heat of fusion. In Table \ref{tableheatoffusion}, the W latent molar heat of fusion is provided as measured by dedicated experiments\,\cite{latfusexp01,latfusexp02,latfusexp03,latfusexp04,latfusexp05,latfusexp06,latfusexp07,latfusexp08,latfusexp09,latfusexp10,latfusexp11,latfusexp12,latfusexp13} or as recommended by theoretical investigations\,\cite{latfusthe01,latfusthe02,latfusthe03} and material handbooks\,\cite{handbooks01,handbooks02,handbooks03,handbooks04,handbooks05}. We point out that the measurement uncertainties in the determination of the heat of fusion with the resistive pulse heating (or dynamic pulse calorimetry) technique are around $10\%$\,\cite{latfusgen01}. Overall, given these uncertainties, the measurements are well clustered around $\Delta{h}_{\mathrm{f}}\simeq50\,$kJ/mol, with $52.3\,$kJ/mol nearly exclusively cited in material handbooks and modelling works.

Some older literature sources recommend a very small value $\Delta{h}_{\mathrm{f}}\simeq35.3\,$kJ/mol, see for instance Refs.\cite{latfusgen00,latfusgen02}, which does not seem to be supported by any measurements. Most probably, this value stems from a semi-empirical relation known as Richard's rule\,\cite{latfusgen03,latfusgen04}; By equating the liquid state with the solid state specific Gibbs free energy $g=h-Ts$ at the melting point, we acquire $\Delta{h}_{\mathrm{f}}=T_{\mathrm{m}}\Delta{s}_{\mathrm{f}}$. Richard's rule is based on positional disorder arguments and empirical observations, it states that the entropy of fusion is a quasi-universal constant for all metals with an approximate value $\Delta{s}_{\mathrm{f}}\simeq{R}$, where $R=N_{\mathrm{A}}k_{\mathrm{B}}$ is the ideal gas constant whose arithmetic value is $R=8.314\,$J/(mol$\cdot$K). This rule allows the calculation of $\Delta{h}_{\mathrm{f}}$ with knowledge of the melting temperature $T_{\mathrm{m}}$ only. For tungsten, we have $T_{\mathrm{m}}=3695\,$K, which translates to $30.72\,$kJ/mol. Modified versions of Richard's rule take into account the average value of the entropy of fusion for bcc and fcc metals, which is $\Delta{s}_{\mathrm{f}}\simeq1.15{R}$\,\cite{latfusthe02} and leads to the inaccurate prediction $35.3\,$kJ/mol. In fact, tungsten is a well-known exception to this entropy rule along with some semi-metals (antimony, bismuth).

\begin{table*}[!ht]
  \centering
  \caption{Tungsten latent molar heat of fusion according to experiments, theoretical investigations and material handbooks.}\label{tableheatoffusion}
  \begin{tabular}{c c c c c c c} \hline
Investigators            & Reference           & \,\,Year\,\, & Value (kJ/mol) & Details  \\    \hline
Lebedev \emph{et al.}    & \cite{latfusexp01}  &    1971      & 54.9           & Experimental (resistive pulse heating)      \\
Martynyuk \emph{et al.}  & \cite{latfusexp02}  &    1975      & 54.4           & Experimental (resistive pulse heating)      \\
Shaner \emph{et al.}     & \cite{latfusexp03}  &    1976      & 46.0           & Experimental (resistive pulse heating)      \\
Seydel \emph{et al.}     & \cite{latfusexp04}  &    1979      & 50.6           & Experimental (resistive pulse heating)      \\
Bonnell                  & \cite{latfusexp05}  &    1983      & 53.0           & Experimental (levitation calorimetry)       \\
Arpaci \& Frohberg       & \cite{latfusexp06}  &    1984      & 50.3           & Experimental (levitation calorimetry)       \\
Berthault \emph{et al.}  & \cite{latfusexp07}  &    1986      & 46.7           & Experimental (resistive pulse heating)      \\
Senchenko \& Sheindlin   & \cite{latfusexp08}  &    1987      & 48.0           & Experimental (resistive pulse heating)      \\
Hixson \& Winkler        & \cite{latfusexp09}  &    1990      & 47.8           & Experimental (resistive pulse heating)      \\
Kaschnitz \emph{et al.}  & \cite{latfusexp10}  &    1990      & 47.1           & Experimental (resistive pulse heating)      \\
McClure \& Cezairliyan   & \cite{latfusexp11}  &    1993      & 48.7           & Experimental (resistive pulse heating)      \\
Pottlacher \emph{et al.} & \cite{latfusexp12}  &    1993      & 52.4           & Experimental (resistive pulse heating)      \\
Kuskova \emph{et al.}    & \cite{latfusexp13}  &    1998      & 45.4           & Experimental (resistive pulse heating)      \\
Gustafson                & \cite{latfusthe01}  &    1985      & 52.3           & Modified experimental input for theory      \\
Grimvall \emph{et al.}   & \cite{latfusthe02}  &    1987      & 52.3           & Modified experimental input for theory      \\
Dinsdale                 & \cite{latfusthe03}  &    1991      & 52.3           & SGTE thermochemical database                \\
Lassner \& Schubert      & \cite{handbooks01}  &    1999      & 46.0           & Handbook of material properties             \\
Lide                     & \cite{handbooks02}  &    2004      & 52.3           & Handbook of material properties             \\
Martienssen \& Warlimont & \cite{handbooks03}  &    2005      & 52.3           & Handbook of material properties             \\
Cardarelli               & \cite{handbooks04}  &    2008      & 52.3           & Handbook of material properties             \\
Shabalin                 & \cite{handbooks05}  &    2014      & 52.3           & Handbook of material properties             \\  \hline \hline
\end{tabular}
\end{table*}

\subsection{The electrical resistivity}\label{resistivity}

\noindent \textbf{Significance.} The role of the electrical resistivity in heat transfer and especially melt motion problems is indirect but can be crucial: (i) It is a key quantity in the determination of the bulk replacement current density $\boldsymbol{J}$, \emph{i.e.} the current that flows through the conductors as a response to thermionic currents emitted or non-ambipolar plasma currents incident at the surface, which leads to a $\boldsymbol{J}\times\boldsymbol{B}$ force density that is believed to drive macroscopic melt layer motion. This is better illustrated by considering the simplest stationary unmagnetized case, where the replacement current is fully described by the steady state continuity equation $\nabla\cdot\boldsymbol{J}=0$ and the electrostatic condition $\nabla\times\boldsymbol{E}=0$\,\cite{EmilspaperN}. For the isotropic tungsten, Ohm's law becomes $\boldsymbol{E}=\rho_{\mathrm{el}}\boldsymbol{J}$ and the irrotational equation can be rewritten as $\rho_{\mathrm{el}}(\nabla\times\boldsymbol{J})+(\nabla\rho_{\mathrm{el}})\times\boldsymbol{J}=0$ or by using the chain rule as $\rho_{\mathrm{el}}(\nabla\times\boldsymbol{J})+(\partial\rho_{\mathrm{el}}/\partial{T})(\nabla{T}\times\boldsymbol{J})=0$. Thus, the temperature dependence of the electrical resistivity is responsible for the second term that can have a significant effect, since sharp temperature gradients are generated by the localized intra-ELM heat fluxes. (ii) It is proportional to the volumetric resistive heating caused by the replacement current that is described by the Joule expression $\rho_{\mathrm{el}}|\boldsymbol{J}|^2$. (iii) For metals, it is inversely proportional to the thermal conductivity, see subsection \ref{conductivity} for details.

\begin{figure*}[!t]
         \centering\lineskip=-4pt
         \subfloat{\includegraphics[width=2.91in]{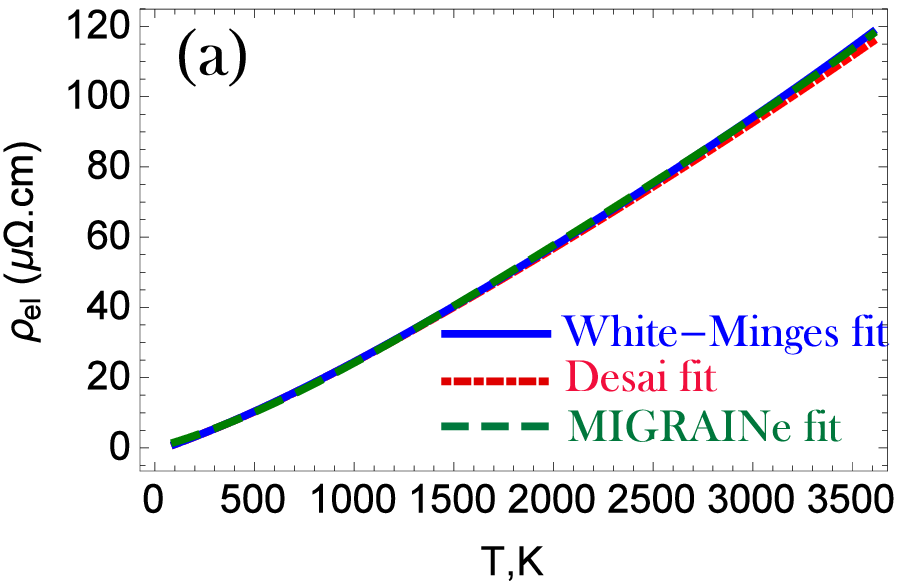}}\qquad
         \subfloat{\includegraphics[width=2.95in]{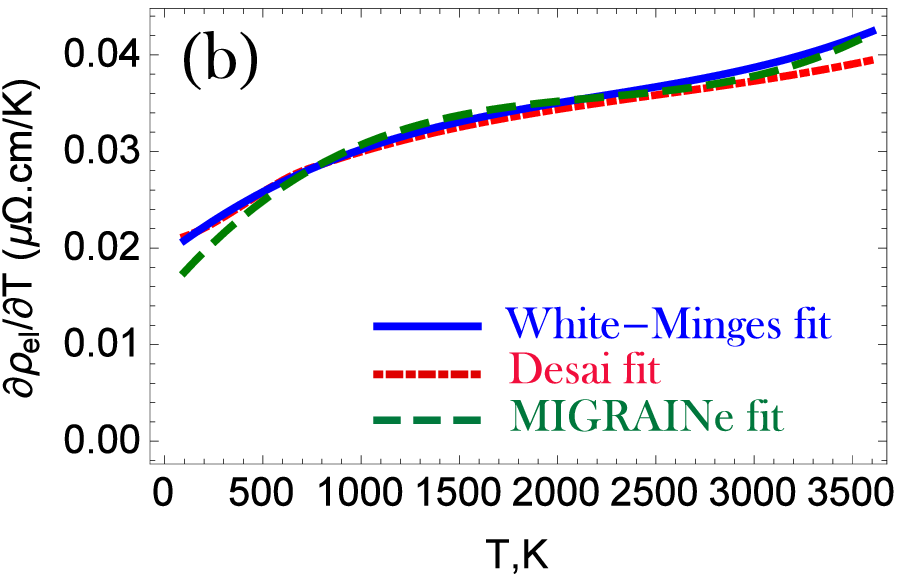}}
\caption{(a) The solid tungsten resistivity as a function of the temperature according to three empirical analytical expressions. (b) The first temperature derivative of the solid tungsten resistivity as a function of the temperature according to these three analytical expressions.}\label{figureWresistivitysolid}
\end{figure*}

\noindent \textbf{Solid tungsten.} In 1984, Desai and collaborators performed the analysis of all $201$ experimental datasets then available for the resistivity of tungsten\,\cite{resistivi01}. A complete dataset covering the temperature range from the neighbourhood of the absolute zero up to $5000\,$K was synthesized from the most reliable measurements over different temperature intervals. For temperatures below the melting point, we shall be completely based on their analysis. In particular, we shall focus on the temperature range $100<T(\mathrm{K})<3695$. For the purpose of numerical manipulation, polynomial expressions were employed to acquire analytical fits for the electrical resistivity. The \emph{Desai fit} reads as\,\cite{resistivi01}
\begin{align*}
\rho_{\mathrm{el}}=
\begin{cases}
+0.000015+7\times10^{-7}T^2+5.2\times10^{-10}T^5\,\,\,\,1\,\mathrm{K}\leq{T}\leq40\,\mathrm{K}\,, \\
+0.14407-1.16651\times10^{-2}T+2.41437\times10^{-4}T^2-3.66335\times10^{-9}T^4\,\,\,\,40\,\mathrm{K}\leq{T}\leq90\,\mathrm{K}\,, \\
-1.06871+2.06884\times10^{-2}T+1.27971\times10^{-6}T^2+8.53101\times10^{-9}T^3-5.14195\times10^{-12}T^4\,\,\,\,90\,\mathrm{K}\leq{T}\leq750\,\mathrm{K}\,,\\
-1.72573+2.14350\times10^{-2}T+5.74811\times10^{-6}T^2-1.13698\times10^{-9}T^3+1.1167\times10^{-13}T^4\,\,\,\,750\,\mathrm{K}\leq{T}\leq3600\,\mathrm{K}\,,
\end{cases}
\end{align*}
where $\rho_{\mathrm{el}}$ is measured in $10^{-8}\,\Omega$m or in $\mu\Omega$cm. Note that the fitting expression proposed by Desai is continuous at its branch points. The following remarks should be explicitly pointed out: \textbf{(i)} The uncertainty in the recommended values employed for the fit is estimated to be $\pm5\%$ below $100\,$K, $\pm3\%$ from $100$ to $300\,$K, $\pm2\%$ from $300$ to $2500\,$K, $\pm3\%$ from $2500$ up to $3600\,$K, $\sim\pm5\%$ in the liquid region. \textbf{(ii)} The recommended polynomial fits do not necessarily imply a recommendation for the temperature derivative of the electrical resistivity. \textbf{(iii)} A large portion of the experimental datasets analyzed by Desai concern mono-crystalline specimens and many times the orientation of the single crystal is not even mentioned. It can be theoretically expected that the resistivity differences between monocrystalline and polycrystalline tungsten are insignificant, because of the bcc tungsten structure. In fact, this has been observed by Desai by inspecting the data. The synthesized Desai dataset was revisited by White and Minges in 1997\,\cite{resistivi02}. These authors fitted a fourth-order polynomial to the - corrected for thermal expansion - recommended values in the range $100<T(\mathrm{K})<3600$. The \emph{White--Minges fit} reads as\,\cite{resistivi02}
\begin{align*}
\rho_{\mathrm{el}}=-0.9680+1.9274\times10^{-2}T+7.8260\times10^{-6}T^2-1.8517\times10^{-9}T^3+2.0790\times10^{-13}T^4\,\,\,\,100\,\mathrm{K}\leq{T}\leq3600\,\mathrm{K}\,,
\end{align*}
where $\rho_{\mathrm{el}}$ is measured in $10^{-8}\,\Omega$m or in $\mu\Omega$cm. This polynomial fit is characterized by a $0.2\%$ rms deviation as well as a maximum deviation of $+0.6\%$ at $150\,$K and $-0.5\%$ at $400\,$K. Finally, in the MIGRAINe dust dynamics code, the resistivity is also an input, since it is needed for the permittivity model that is employed in the Mie calculation of the emissivity\,\cite{introduct12}. A polynomial fit has been employed in the MIGRAINe code that is similar to the White--Minges expression. The \emph{MIGRAINe fit} reads as\,\cite{resistivi03}
\begin{align*}
\rho_{\mathrm{el}}=+0.000015+1.52\times10^{-2}T+1.2003\times10^{-5}T^2-3.3467\times10^{-9}T^3+3.7906\times10^{-13}T^4\,\,\,\,T\leq3600\,\mathrm{K}\,,
\end{align*}
where again $\rho_{\mathrm{el}}$ is measured in $10^{-8}\,\Omega$m or in $\mu\Omega$cm. A comparison between the resistivities and the resistivity temperature derivatives stemming from the three different fits can be found in figure \ref{figureWresistivitysolid}. The deviations between the different fits are very small also for the temperature derivative. It is preferable though that the White--Minges fit is employed in future applications and extrapolated up to the actual melting point of $3695\,$K. The justification for the choice of this fit will be provided in the following paragraph.

\noindent \textbf{Discontinuity at the melting point.} The electrical resistivity of all elements has a discontinuity at the melting point. For most liquid metals $\rho^{\mathrm{l}}_{\mathrm{el}}>\rho^{\mathrm{s}}_{\mathrm{el}}$ but there are few exceptions\,\cite{liquidsbook}. Reliable measurements of the W electrical resistivity asymptotically before and after the melting point have been outlined in Table \ref{tableresistivitydiscontinuity}. Their mean values are $\langle\rho^{\mathrm{s}}_{\mathrm{el}}\rangle\simeq121\,\mu\Omega$cm and $\langle\rho^{\mathrm{l}}_{\mathrm{el}}\rangle\simeq136\,\mu\Omega$cm. They are very close to the measurements of Seydel \& Fucke\,\cite{resistivi04}, whose measurements we are going to adopt not only for the discontinuity but also for the liquid state. The extrapolated value of the Desai fit is $\rho^{\mathrm{s}}_{\mathrm{el}}\simeq119\,\mu\Omega$cm, the extrapolated value of the White-Minges fit is $\rho^{\mathrm{s}}_{\mathrm{el}}\simeq122\,\mu\Omega$cm and the extrapolated value of the MIGRAINe fit is $\rho^{\mathrm{s}}_{\mathrm{el}}\simeq122\,\mu\Omega$cm. Therefore, we conclude that the White--Minges fit (but also the MIGRAINe fit) can be extrapolated from $3600\,$K to $3695\,$K with a negligible error.

\begin{table*}[!b]
  \centering
  \caption{The W electrical resistivity at the solid-liquid phase transition; the values from the solid side $\rho^{\mathrm{s}}_{\mathrm{el}}=\rho_{\mathrm{el}}(T_{\mathrm{m}}^-)$
   and the liquid side $\rho^{\mathrm{l}}_{\mathrm{el}}=\rho_{\mathrm{el}}(T_{\mathrm{m}}^+)$, as well as the discontinuity magnitude $\Delta\rho_{\mathrm{el}}=\rho^{\mathrm{l}}_{\mathrm{el}}-\rho^{\mathrm{s}}_{\mathrm{el}}$. The first two datasets\,\cite{latfusexp01,latfusexp02} have been corrected for thermal expansion effects following Ref.\cite{resistivi01}.}\label{tableresistivitydiscontinuity}
  \begin{tabular}{c c c c c c c c} \hline
Investigators            & Reference          & \,\,Year\,\, & \,\,$\rho^{\mathrm{s}}_{\mathrm{el}}$ ($\mu\Omega$cm)\,\, & \,\,$\rho^{\mathrm{l}}_{\mathrm{el}}$ ($\mu\Omega$cm)\,\, & \,\,$\Delta\rho_{\mathrm{el}}$ ($\mu\Omega$cm) \\ \hline
Lebedev \emph{et al.}    & \cite{latfusexp01} &    1971      & 121                                 & 127                                 & 6                            \\
Martynyuk \emph{et al.}  & \cite{latfusexp02} &    1975      & 118                                 & 125                                 & 7                            \\
Shaner \emph{et al.}     & \cite{latfusexp03} &    1976      & 118                                 & 132                                 & 14                           \\
Seydel \emph{et al.}     & \cite{latfusexp04} &    1979      & 120                                 & 137                                 & 17                           \\
Seydel \& Fucke          & \cite{resistivi04} &    1980      & 121                                 & 135                                 & 14                           \\
Desai \emph{et al.}      & \cite{resistivi01} &    1984      & 121                                 & 131                                 & 10                           \\
Berthault \emph{et al.}  & \cite{latfusexp07} &    1986      & 123                                 & 138                                 & 15                           \\
Hixson \& Winkler        & \cite{latfusexp09} &    1990      & 126                                 & 146                                 & 20                           \\
Kaschnitz \emph{et al.}  & \cite{latfusexp10} &    1990      & 118                                 & 138                                 & 20                           \\
Pottlacher \emph{et al.} & \cite{latfusexp12} &    1993      & 126                                 & 145                                 & 19                           \\
Kuskova \emph{et al.}    & \cite{latfusexp13} &    1998      & 120                                 & 140                                 & 20                           \\ \hline \hline
\end{tabular}
\end{table*}

\begin{figure*}[!ht]
         \centering\lineskip=-4pt
         \subfloat{\includegraphics[width=2.80in]{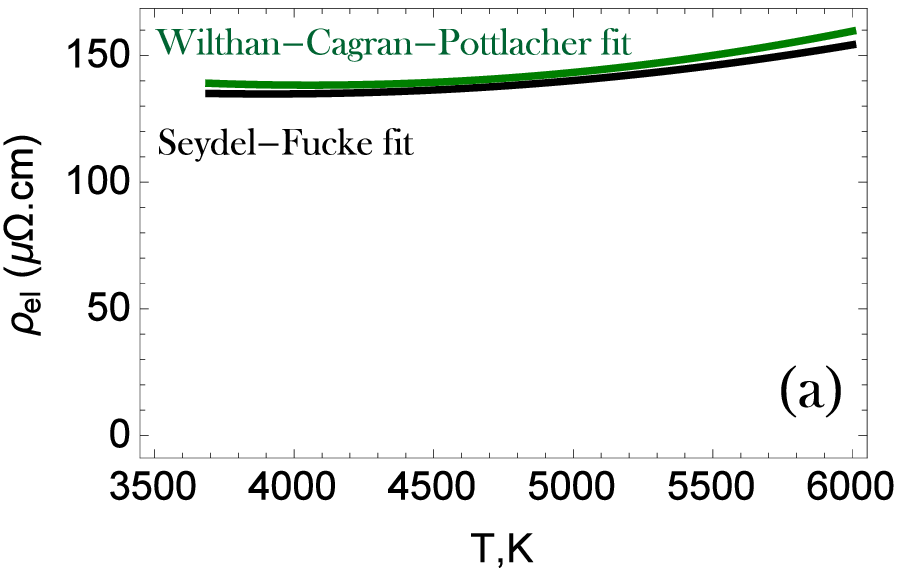}}\qquad
         \subfloat{\includegraphics[width=2.98in]{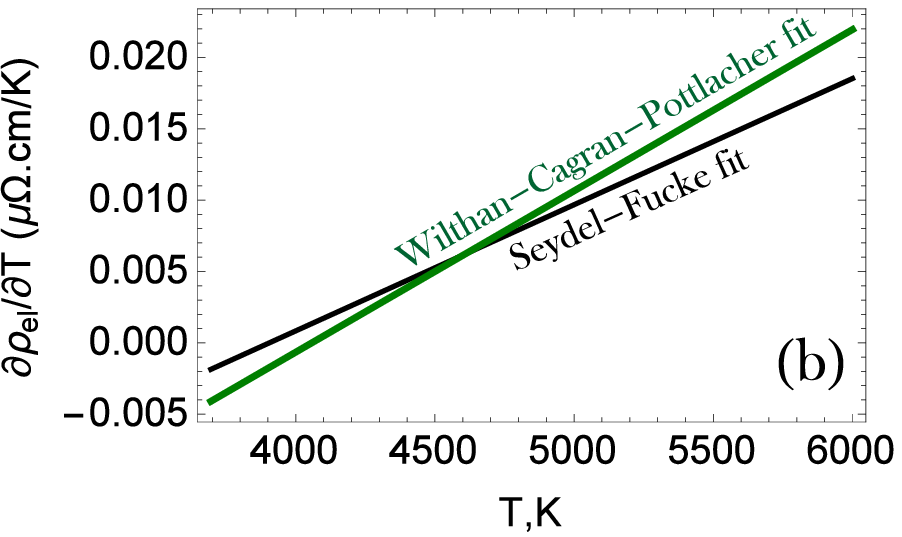}}
\caption{(a,b) The W electrical resistivity and its first temperature derivative as a function of the temperature across the liquid state according to two empirical analytical expressions.}\label{figureWresistivityliquid}
\end{figure*}

\noindent \textbf{Liquid tungsten.} The electrical resistivity of elemental liquid metals generally exhibits two tendencies\,\cite{resistivi01,liquidsbook,resistivi04}: \textbf{(i)} a monotonous increase beyond the melting point at a much slower pace than the solid state increase (e.g. refractory metals such as Ti, V, Mo), \textbf{(ii)} a very slow decrease right after the melting point followed by an increase again at a much slower pace than the solid state increase (e.g. the low melting point Zn). Tungsten belongs to the second group\,\cite{resistivi04}. The experimental results have been fitted with a second-order polynomial. The \emph{Seydel--Fucke fit} reads as
\begin{equation*}
\rho_{\mathrm{el}}=135-1.855\times10^{-3}(T-T_{\mathrm{m}})+4.420\times10^{-6}(T-T_{\mathrm{m}})^2\,\,\,\,T\geq3695\,\mathrm{K}\,,
\end{equation*}
where $\rho_{\mathrm{el}}$ is measured in $10^{-8}\,\Omega$m. We point out that there are some uncertainties in the temperature measurements due to the lack of data for the temperature dependence of the liquid tungsten emissivity. A constant emissivity has been assumed across the liquid phase, which can be expected to translate from a $5\%$ $T$-uncertainty near the melting point to a $10\%$ $T$-uncertainty close to $6000\,$K. On the other hand, the uncertainties in the resistivity measurements should be $5-6\%$. To our knowledge, the only alternative analytical expression for the resistivity of liquid tungsten has been provided by Wilthan \emph{et al.}\,\cite{resistivi05}, see also Refs.\cite{resistivi06,resistivi07}. The experiments were performed from $423\,$K to $5400\,$K and a polynomial fit was employed (including expansion effects). The \emph{Wilthan--Cagran--Pottlacher fit} reads as\,\cite{resistivi05,resistivi06}
\begin{align*}
\rho_{\mathrm{el}}=231.3-4.585\times10^{-2}T+5.650\times10^{-6}T^2\,\,\,\,T\geq3695\,\mathrm{K}\,,
\end{align*}
where again $\rho_{\mathrm{el}}$ is measured in $10^{-8}\,\Omega$m or in $\mu\Omega$cm. From figure \ref{figureWresistivityliquid}, it is evident that the analytical fits are nearly identical.

\begin{figure*}[!ht]
         \centering\lineskip=-4pt
         \subfloat{\includegraphics[width=2.90in]{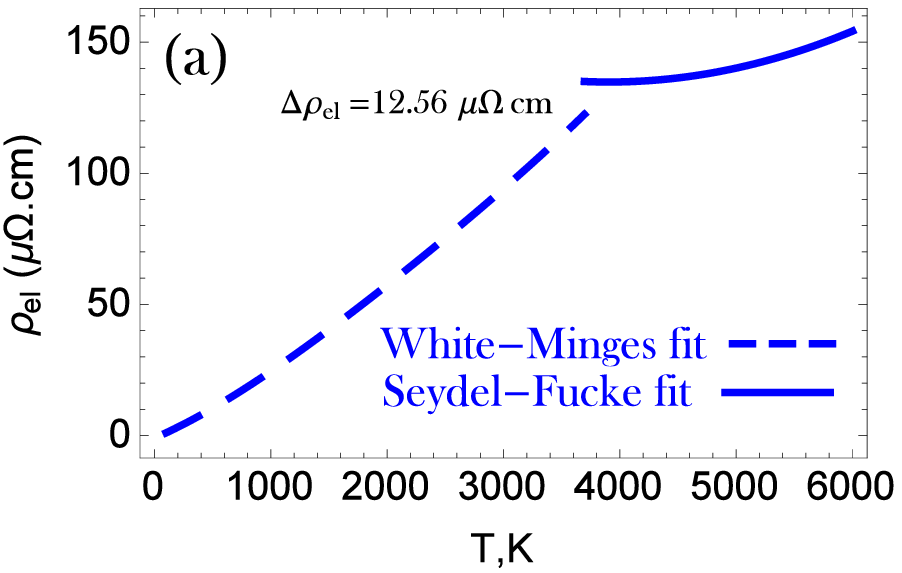}}\qquad
         \subfloat{\includegraphics[width=2.95in]{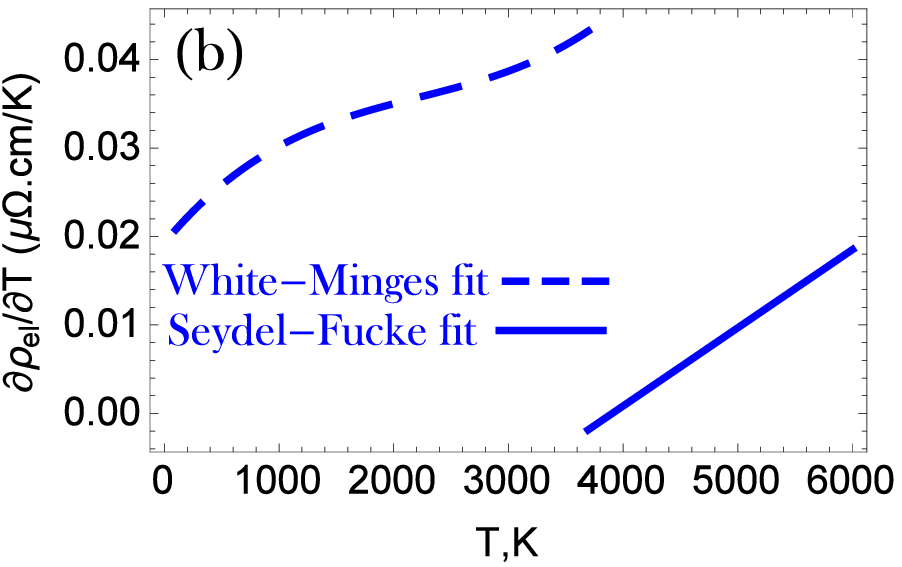}}
\caption{(a,b) The recommended treatment for the W electrical resistivity and its first temperature derivative for temperatures ranging from $100$ up to $6000\,$K.}\label{figureWresistivityfull}
\end{figure*}

\noindent \textbf{Recommended description.} The analytical description of the W electrical resistivity consists of employing the White--Minges fit in the temperature range $100<T(\mathrm{K})<3695$ and the Seydel--Fucke fit in the temperature range $3695<T(\mathrm{K})<6000$. The W electrical resistivity is illustrated in figure \ref{figureWresistivityfull}. It is worth pointing out that the relative magnitude of the discontinuity of the resistivity at the liquid-solid phase transition is very small, whereas the relative magnitude of the discontinuity of the resistivity temperature derivative is very large (notice also the sign reversal). Finally, we also note that, in the ITER database, a cubic polynomial expression is recommended for the temperature range from $300$ to $3300\,$K\,\cite{introduct05}. This expression is nearly identical to the Desai, White--Minges and MIGRAINe fits for solid tungsten.

\subsection{The specific isobaric heat capacity}\label{capacity}

\begin{figure*}[!ht]
         \centering\lineskip=-4pt
         \subfloat{\includegraphics[width=2.80in]{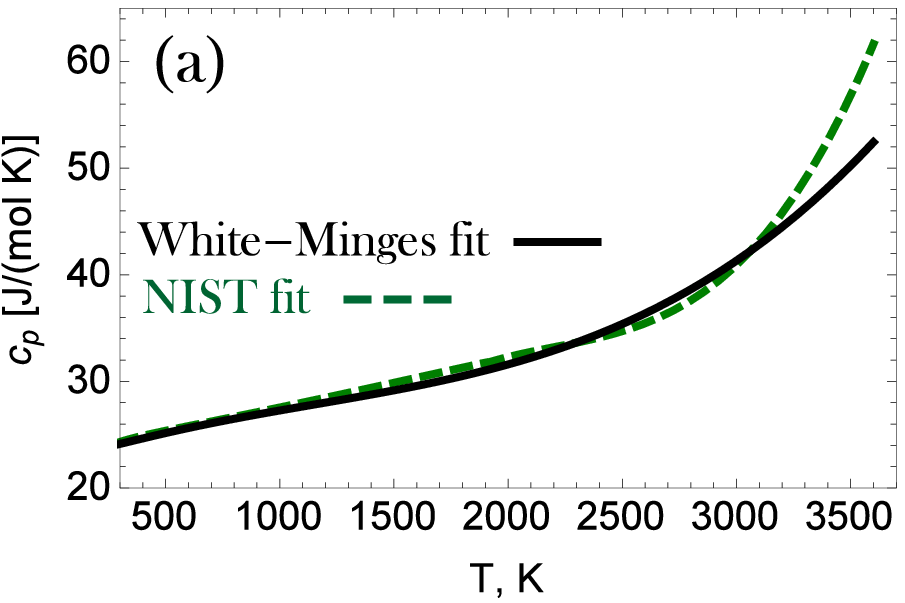}}\qquad
         \subfloat{\includegraphics[width=2.93in]{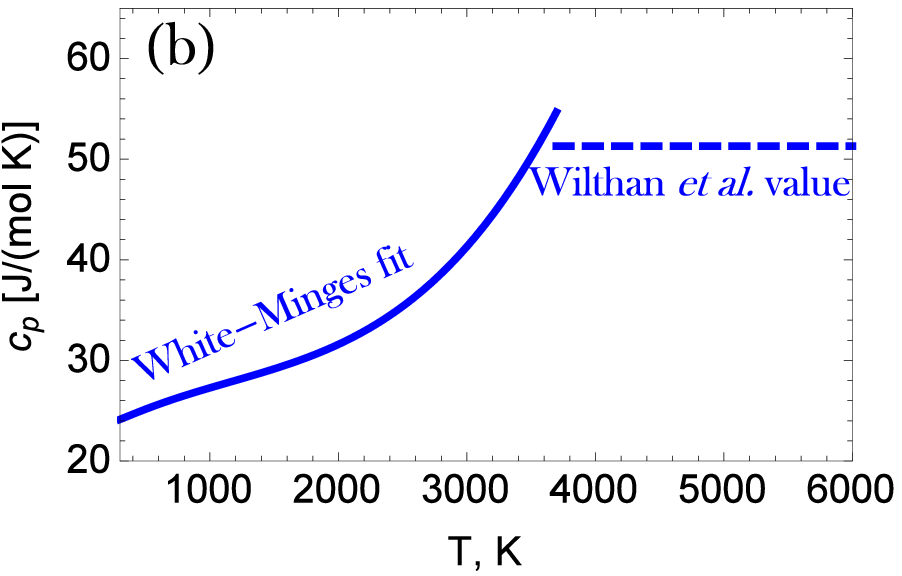}}
\caption{(a) The W specific isobaric heat capacity in the solid state as a function of the temperature according to two empirical analytical expressions. (b) A complete analytical description of the W specific isobaric heat capacity from $300$ to $6000\,$K by employing the White--Minges fit in the range $300<T(\mathrm{K})<3695$ and the constant value of Wilthan \emph{et al.} $c_{\mathrm{p}}=51.3\,$J/(mol\,K) in the range $3695<T(\mathrm{K})<6000$.}\label{figureWcapacity}
\end{figure*}

\noindent \textbf{Solid tungsten.} The fourth and last edition of the NIST-JANAF Thermochemical Tables was published in 1998, but the tungsten data were last reviewed in June 1966\,\cite{heatcapaci1}. Measurements from 12 different sources were employed that were published from 1924 up to 1964. Only four of these datasets extend at temperatures beyond $2000\,$K, whereas five datasets are exclusively focused below the room temperature. The NIST webpage provides a Shomate equation fit in the temperature intervals $298<T(\mathrm{K})<1900$ and $1900<T(\mathrm{K})<3680$. The \emph{NIST fit} reads as\,\cite{heatcapaci2}
\begin{align*}
c_{\mathrm{p}}=
\begin{cases}
+23.9593+2.63968\times10^{-3}T+1.25775\times10^{-6}T^2-2.54642\times10^{-10}T^3-\displaystyle\frac{4.8407\times10^{4}}{T^{2}}\,\,\,\,298\,\mathrm{K}\leq{T}\leq1900\,\mathrm{K}\,, \\
-22.5764+9.02798\times10^{-2}T-4.42715\times10^{-5}T^2+7.17663\times10^{-9}T^3-\displaystyle\frac{2.40974\times10^{7}}{T^{2}}\,\,\,\,1900\,\mathrm{K}\leq{T}\leq3680\,\mathrm{K}\,,
\end{cases}
\end{align*}
where $c_{\mathrm{p}}$ is measured in J/(mol\,K). In 1997, White and Minges\,\cite{resistivi02} revisited an earlier synthetic dataset of recommended values\,\cite{heatcapaci3}. In the range above the room temperature, eleven datasets (dating up to 1994) were selected. The \emph{White--Minges fit} reads as\,\cite{resistivi02}
\begin{align*}
c_{\mathrm{p}}=21.868372+8.068661\times10^{-3}T-3.756196\times10^{-6}T^2+1.075862\times10^{-9}T^3+\displaystyle\frac{1.406637\times10^{4}}{T^2}\,\,\,\,300\,\mathrm{K}\leq{T}\leq3400\,\mathrm{K}\,,
\end{align*}
where $c_{\mathrm{p}}$ is measured in J/(mol\,K). This fit is characterized by a $1.1\%$ rms deviation, the deviation from the mean is generally less than $1\%$ below $1000\,$K and less than $2.5\%$ above $1000\,$K. We point out that there are two misprints in the fitting expression as quoted in the original work\,\cite{resistivi02}. As illustrated in figure \ref{figureWcapacity}a, the two expressions begin to strongly diverge above $2900\,$K; the high temperature measurements employed in the NIST fit are far less reliable.

\noindent \textbf{Liquid tungsten.} Measurements on free-electron-like elemental metals with low melting points\,\cite{latfusgen03} as well as recent experiments on elemental transition metals\,\cite{latfusexp12,resistivi07} indicate that the enthalpy of liquid metals increases nearly linearly with the temperature over a wide range. In the case of liquid tungsten, the literature consensus is also that the enthalpy at constant pressure is a linear function of the temperature. This implies a constant isobaric heat capacity, courtesy of $(\partial{H}/\partial{T})_{P}=C_{\mathrm{p}}$. Thus, also the specific isobaric heat capacity $c_{\mathrm{p}}=\partial{C}_{\mathrm{p}}/\partial{m}$ should be constant. However, there is a disagreement concerning the exact value: \textbf{(i)} The NIST-JANAF recommended value is $c_{\mathrm{p}}=35.564\,$J/(mol\,K). It is very outdated, being based on experiments that were carried out prior to 1961, \emph{i.e.} many years before the dynamic pulse calorimetry or levitation calorimetry methods were developed. Unfortunately, this value is quoted in material property handbooks\,\cite{handbooks01}. \textbf{(ii)} More reliable measurements provide values that are clustered around $c_{\mathrm{p}}=52\,$J/(mol\,K). We have $c_{\mathrm{p}}=51.8\,$J/(mol\,K)\,\cite{latfusexp03}, $c_{\mathrm{p}}=57.0\,$J/(mol\,K)\,\cite{latfusexp04}, $c_{\mathrm{p}}=55.1\,$J/(mol\,K)\,\cite{latfusexp07}, $c_{\mathrm{p}}=48.2\,$J/(mol\,K)\,\cite{latfusexp09}, $c_{\mathrm{p}}=56.1\,$J/(mol\,K)\,\cite{latfusexp10}, $c_{\mathrm{p}}=52.9\,$J/(mol\,K)\,\cite{latfusexp12}, $c_{\mathrm{p}}=53.7\,$J/(mol\,K)\,\cite{latfusexp13}. Such deviations are justified in view of the fact that $c_{\mathrm{p}}$ is not directly obtained by the measurements but after post-processing (graphical determination from the slope of the enthalpy versus the temperature trace) and thus is subject to an uncertainty of around $10\%$\,\cite{latfusgen01}. \textbf{(iii)} To our knowledge, the most contemporary experiments are those performed by Wilthan \emph{et al.}\,\cite{resistivi05} in 2005, who performed measurements up to $5400\,$K and found a constant liquid W value $c_{\mathrm{p}}=51.3\,$J/(mol\,K), that we shall adopt.

\begin{figure*}[!ht]
         \centering\lineskip=-4pt
         \subfloat{\includegraphics[width=2.90in]{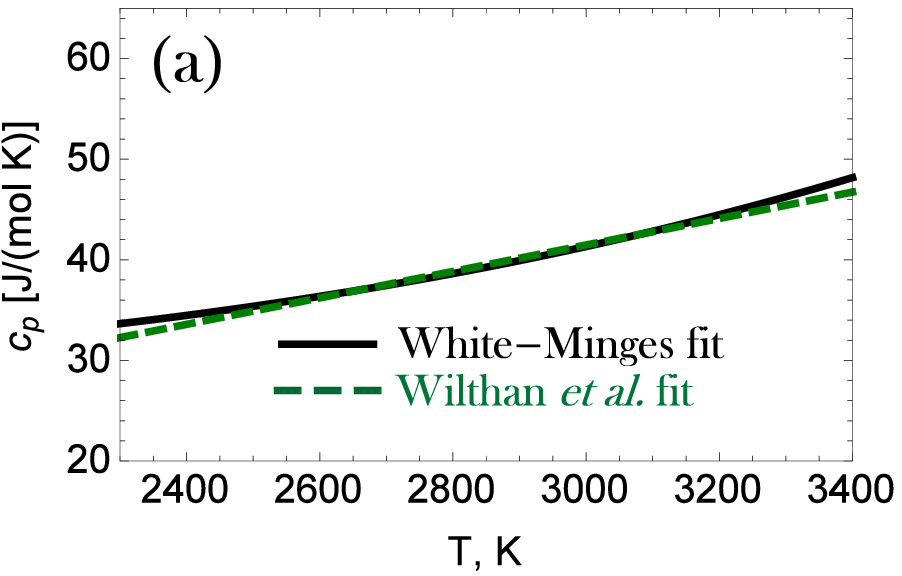}}\qquad
         \subfloat{\includegraphics[width=2.90in]{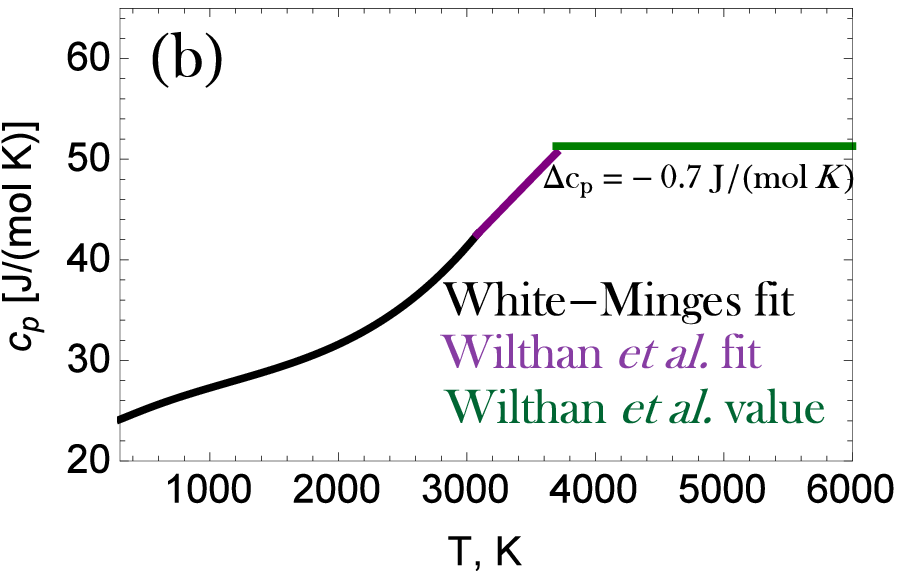}}
\caption{(a) The specific isobaric heat capacity of solid state tungsten in the range $2300<T(\mathrm{K})<3400$ according to two empirical analytical expressions. (b) The complete recommended analytical description of the W specific isobaric heat capacity from $300$ to $6000\,$K.}\label{figureWcapacityfull}
\end{figure*}

\noindent \textbf{Recommended description.} \textbf{(i)} A complete analytical description of the tungsten specific isobaric heat capacity can be constructed by combining the White--Minges fit in the temperature range $300<T(\mathrm{K})<3695$ and the constant value $c_{\mathrm{p}}=51.3\,$J/(mol\,K) in the temperature range $3695<T(\mathrm{K})<6000$. See also figure \ref{figureWcapacity}b. This implies that the White--Minges fit needs to be extrapolated in the temperature range $3400<T(\mathrm{K})<3695$. This leads to $c_{\mathrm{p}}^{\mathrm{s}}\simeq54.7\,$J/(mol\,K) and thus to $\Delta{c}_{\mathrm{p}}\simeq3.4\,$J/(mol\,K). However, as can be observed in figure \ref{figureWcapacity}a, the heat capacity starts rapidly increasing at high temperatures, which implies that any extrapolation can lead to significant errors. \textbf{(ii)} Wilthan \emph{et al.} have also provided an analytical fit for the tungsten specific enthalpy in the range $2300<T(\mathrm{K})<3687$\,\cite{resistivi05}. Their fit reads as $h(T)=83.342+0.011T+3.576\times10^{-5}T^2\,$(kJ/kg). It would be certainly preferable that the heat capacity was calculated from the local slopes of the experimental data, but here we have to differentiate the above fitting expression, which yields $c_{\mathrm{p}}=11+7.152\times10^{-2}T\,$[J/(kg\,K)] or $c_{\mathrm{p}}=2.022+1.315\times10^{-2}T\,$[J/(mol\,K)]. Therefore, we have $c_{\mathrm{p}}^{\mathrm{s}}\simeq50.6\,$J/(mol\,K) and thus $\Delta{c}_{\mathrm{p}}\simeq-0.7\,$J/(mol\,K). \textbf{(iii)} Both results are physically acceptable; At the melting point, the difference in the heat capacity of metals between the solid and the liquid phases is rather small and it can be of either sign\,\cite{latfusgen03,handbooks06}. \textbf{(iv)} In their common range of validity, the fits agree exceptionally well, see figure \ref{figureWcapacityfull}a, but they start diverging at both the interval endpoints. It is preferable to avoid any extrapolations and employ both fits. We shall first calculate their highest temperature intersection point, which is $T\simeq3080\,$K. This allows us to connect the two fitting expressions in a continuous manner. The recommended description has the form
\begin{align*}
c_{\mathrm{p}}=
\begin{cases}
21.868372+8.068661\times10^{-3}T-3.756196\times10^{-6}T^2+1.075862\times10^{-9}T^3+\displaystyle\frac{1.406637\times10^{4}}{T^2}\,\,\,\,300\,\mathrm{K}\leq{T}\leq3080\,\mathrm{K} \\
2.022+1.315\times10^{-2}T\,\,\,\,3080\,\mathrm{K}\leq{T}\leq3695\,\mathrm{K} \\
51.3\,\,\,\,{T}\geq3695\,\mathrm{K}
\end{cases}
\end{align*}
where $c_{\mathrm{p}}$ is again measured in J/(mol\,K). The recommended analytical description is illustrated in figure \ref{figureWcapacityfull}b.

\begin{figure*}[!ht]
         \centering\lineskip=-4pt
         \subfloat{\includegraphics[width=2.90in]{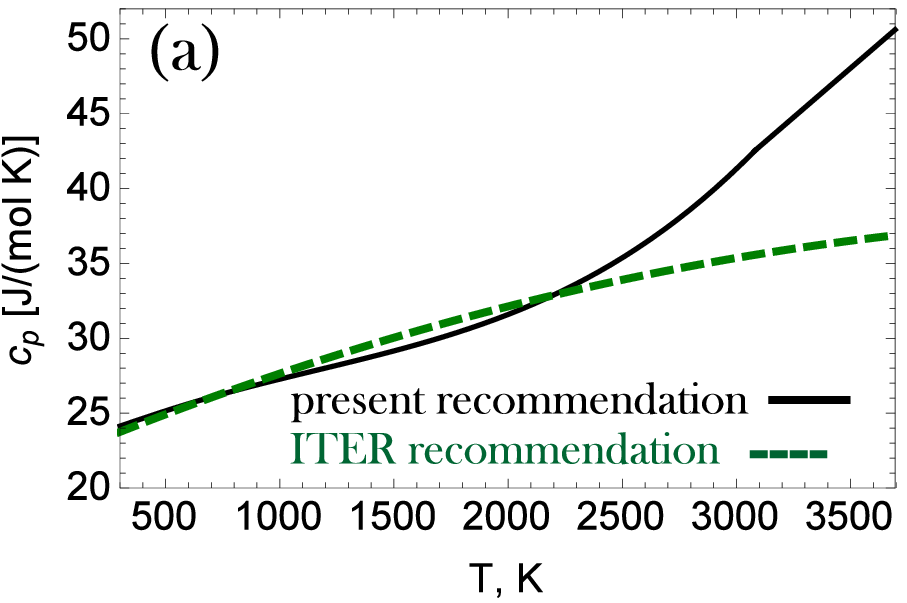}}\qquad
         \subfloat{\includegraphics[width=2.99in]{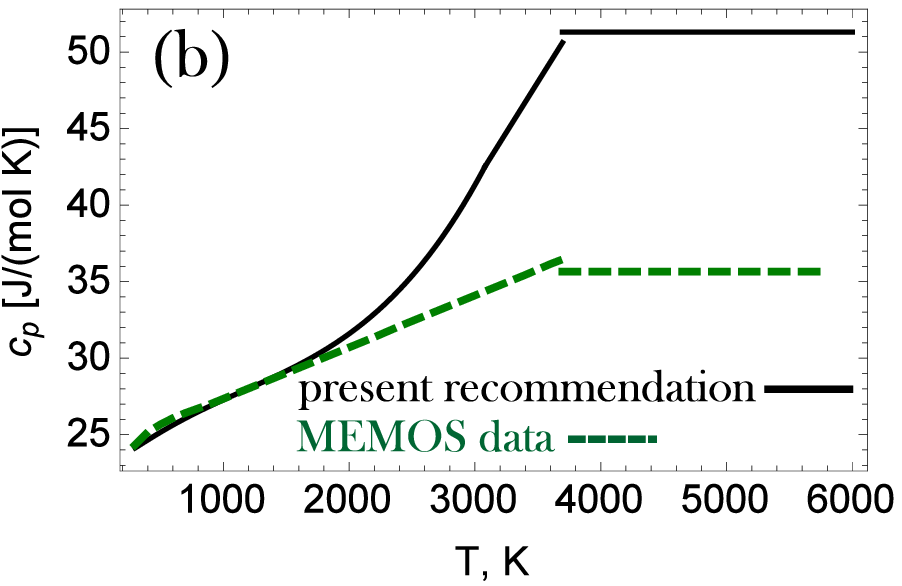}}
\caption{(a) Comparison of the recommended analytical description of the W specific isobaric heat capacity with the ITER database recommendation in the temperature range from $300$ to $3695\,$K. (b) Comparison of the recommended analytical description of the W specific isobaric heat capacity with the MEMOS code description in the temperature range from $300$ to $6000\,$K.}\label{figureWcapacitycomp}
\end{figure*}

\noindent \textbf{Comparison with the fusion literature}. \emph{In the ITER database}; a quadratic polynomial expression is recommended which is valid in the range $273-3100\,$K\,\cite{introduct05}. It originates from fitting to a synthetic dataset, whose high temperature part is heavily based on measurements provided in the classic 1971 compendium by Touloukian\,\cite{TouloukianB}. As illustrated in figure \ref{figureWcapacitycomp}a, the ITER recommendation is outdated. The underestimations of the heat capacity start from $2200\,$K and monotonically increase up to $3100\,$K. In the extrapolated range $3100-3695\,$K, the situation becomes progressively worse with the underestimation reaching $40\%$ close to the phase transition. \emph{In the MEMOS code}; a dataset based on Touloukian's compendium is implemented for interpolations in the solid state, whereas the constant NIST-JANAF value of $c_{\mathrm{p}}=35.564\,$J/(mol\,K) is employed for the liquid state\,\cite{introduct08}. As evident from figure \ref{figureWcapacitycomp}b, in MEMOS, the heat capacity is underestimated from $1700\,$K with the deviations approaching $\sim40\%$ from above $\sim3000\,$K and across the entire liquid state. \emph{Consequences}; Underestimation of the heat capacity translates to overestimation of the temperature in the MEMOS simulations compared to the experiments, which could be erroneously attributed to a decreased heat flux incidence from the inter-ELM and intra-ELM plasma. Furthermore, this implies an overestimation of the melt layer depth and a premature initiation of bulk melting during prolonged exposures.

\subsection{The thermal conductivity}\label{conductivity}

\noindent \textbf{Preliminaries.} \textbf{(i)} In condensed matter, heat transfer is mediated by the collisional transport of valence electrons and lattice waves. In metals, the electron contribution dominates over the phonon contribution (which is limited by Umklapp processes) with the exception of samples with high impurity concentration at very low temperatures\,\cite{conductsol4,conductliq4}. Due to the fact that the valence electrons are responsible for both charge and heat transfer in metals, a proportionality between the thermal conductivity and the electrical conductivity can be expected. This is expressed by the so-called Wiedemann-Franz law that can be derived within Sommerfeld's free-electron theory and the Lorentz gas approximation, it reads as $k=\left[(\pi^2k_{\mathrm{b}}^2)/(3e^2)\right][T/\rho_{\mathrm{el}}]$\,\cite{conductsol4,conductliq4,conductliq3}. The term in brackets is known as the Lorenz number and its nominal value is $L_0=2.443\times10^{-8}\,\mathrm{W}\Omega\mathrm{K}^{-2}$. \textbf{(ii)} The resistance to heat transfer by electrons originates from collisions with phonons and collisions with atomic impurities, crystal boundaries, lattice imperfections. The coupling between these collisional contributions is limited, which implies an additivity that is expressed by Matthiessen's rule. However, with the exception of extreme cases where the impurity/imperfection concentration is very large, electron-phonon collisions dominate already from the room temperature\,\cite{conductliq4}. Thus, for our temperature range of interest, the thermal conductivity can be expected to be weakly dependent on crystalline structure details.

\begin{figure*}[!ht]
         \centering\lineskip=-4pt
         \subfloat{\includegraphics[width=2.92in]{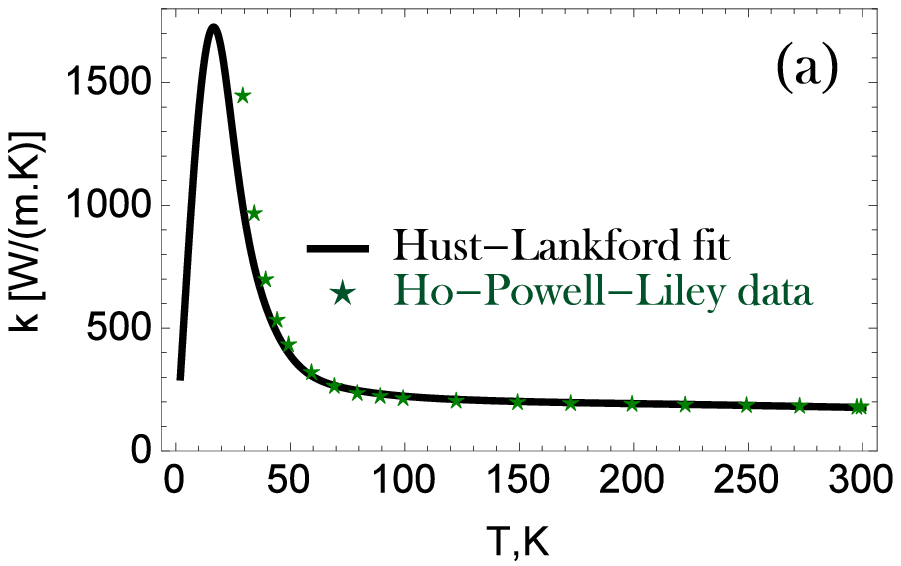}}\qquad
         \subfloat{\includegraphics[width=2.90in]{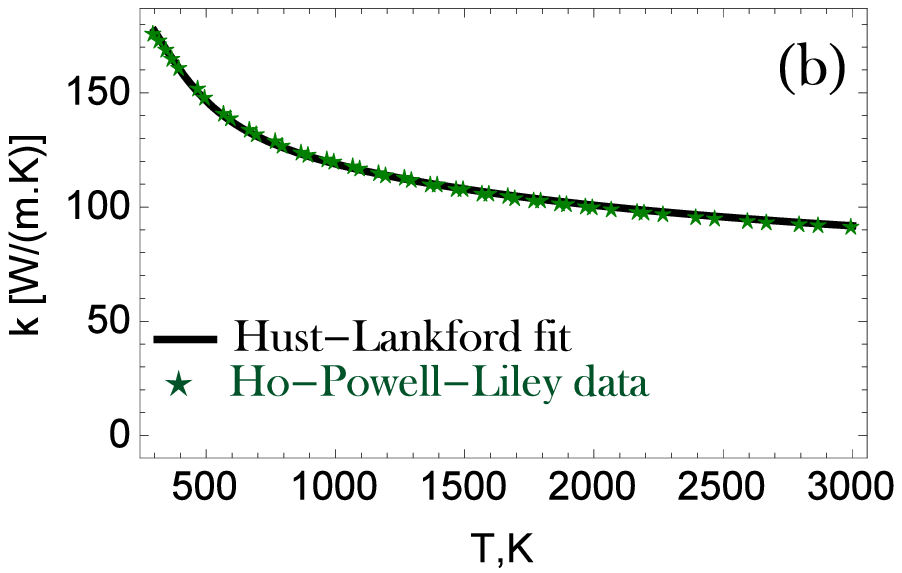}}
\caption{The solid tungsten thermal conductivity. Comparison of the Hust--Lankford analytical fit\,\cite{conductsol3} with the Ho--Powell--Liley\,\cite{conductsol1,conductsol2} recommended dataset in the (a) low temperature interval $2<T(\mathrm{K})<300$, where the deviations rapidly increase as $T<40\,$K, (b) intermediate and high temperature interval $300<T(\mathrm{K})<3000$, where the agreement is excellent.}\label{figureWthermalconductivitysolid}
\end{figure*}

\noindent \textbf{Solid tungsten.} \emph{In 1972}; Ho, Powell and Liley provided recommended and estimated thermal conductivity values for all elements with atomic numbers up to $Z=105$\,\cite{conductsol1,conductsol2}. These recommended datasets were synthesized for $82$ elements after the careful analysis of $5200$ different sets of experimental measurements. Their recommended dataset for tungsten will not be employed for the determination of the fitting expression but will be employed for comparison with our recommended treatment. \emph{In 1984}; Hust and Lankford critically analyzed all literature data on the thermal conductivity of four reference metals (aluminium, copper, iron, tungsten) for temperatures up to melting as well as provided analytical fits based on theoretical descriptions\,\cite{conductsol3}. Their analysis was later closely followed by White and Minges\,\cite{resistivi02}. Intricate details of their analysis and, in particular, their utilization of the residual resistivity ratio will not be discussed here, since they are important for the low temperature part of the thermal conductivity ($\lesssim100\,$K), which is not relevant for fusion applications. They utilized $13$ datasets for their fit, which contain experimental results from $2\,$K up to $3000\,$K (only four datasets contained measurements above $2000\,$K).  The basic ingredients of the \emph{Hust--Lankford fit} for tungsten are the electron-defect interaction term $W_{\mathrm{o}}$ ($\propto{T}^{-1}$), the electron-phonon interaction term $W_{\mathrm{i}}$ (approximately $\propto{T}^2$), the interaction coupling term $W_{\mathrm{io}}$ (nearly zero for tungsten) and the mathematical residual deviation term $W_{\mathrm{c}}$. These terms are combined to provide the thermal conductivity in a manner reminiscent of Matthiessen's rule. The analytical expressions and their connection to the thermal conductivity read as\,\cite{conductsol3}
\begin{align*}
&W_{\mathrm{c}}(T)=-0.00085\ln{\left(\frac{T}{130}\right)}\exp{\left\{-\left[\ln{\left(\frac{T}{230}\right)}\frac{1}{0.7}\right]^2\right\}}+0.00015\exp{\left\{-\left[\ln{\left(\frac{T}{3500}\right)}\frac{1}{0.8}\right]^2\right\}}\,\\
&\qquad\,\,\,\,\,\,\,\,\,\,\,\,+0.0006\ln{\left(\frac{T}{90}\right)}\exp{\left\{-\left[\ln{\left(\frac{T}{80}\right)}\frac{1}{0.4}\right]^2\right\}}+0.0003\ln{\left(\frac{T}{24}\right)}\exp{\left\{-\left[\ln{\left(\frac{T}{33}\right)}\frac{1}{0.5}\right]^2\right\}}\,,\\
&W_{\mathrm{i}}(T)=\frac{P_1T^{P_2}}{1+P_1P_3T^{(P_2+P_4)}\exp{\left[-\left({P_5}/{T}\right)^{P_6}\right]}}\,,\quad\,W_{\mathrm{o}}(T)=\frac{\beta}{T}\,,\quad\,k=\frac{1}{W_{\mathrm{o}}(T)+W_{\mathrm{i}}(T)+W_{\mathrm{c}}(T)}\,.
\end{align*}
The constant $\beta$ has been chosen to correspond to a residual resistivity ratio of $300$, whereas the $P_i$ parameters were determined by least square fits of the combined dataset. Their arithmetic values are\,\cite{conductsol3}
\begin{align*}
\beta=0.006626\,,\quad\,P_1=31.70\times10^{-8}\,,\quad\,P_2=2.29\,,\quad\,P_3=541.3\,,\quad\,P_4=-0.22\,,\quad\,P_5=69.94\,,\quad\,P_6=3.557\,.
\end{align*}
Surpringly, a comparison of the fit with the tabulated values reveals deviations below $90\,$K. This can either originate from misprints in the residual deviation $W_{\mathrm{c}}$ or from improper rounding-off of the least square coefficients. Since these deviations lie well below our temperature range of interest, we have not pursued this issue further. For completeness, the functional form of the tungsten thermal conductivity according to the Hust--Lankford fit is illustrated in figure \ref{figureWthermalconductivitysolid}. The plot covers the full temperature range of validity, $2<T(\mathrm{K})<3000$, but the fit will only be utilized in the temperature range $300<T(\mathrm{K})<3000$. In the latter range, the comparison with the Ho--Powell--Liley recommended dataset reveals a remarkable agreement. On the other hand, in the low temperature range, there are very strong deviations below $40\,$K (exceeding by far the selected plot scale). The emergence of these deviations is theoretically expected; they are a direct consequence of the electron-defect interaction term, which becomes dominant at very low temperatures and is a very sensitive function of the sample purity\,\cite{conductsol4}. The Hust--Lankford fitting function is relatively cumbersome for numerical simulations. Its complexity stems from the low temperature maximum of the thermal conductivity, whose position lies well below fusion regimes of interest. An alternative empirical expression has been found by digitizing the Hust--Lankford fitting function with sampling steps of $50\,$K from $300\,$K to $3700\,$K and by least squares fitting the emerging dataset to the Shomate equation. This \emph{modified Hust--Lankford fit} reads as
\begin{align*}
k=149.441-45.466\times10^{-3}T+13.193\times10^{-6}T^2-1.484\times10^{-9}T^3+\frac{3.866\times10^{6}}{T^2}\,,
\end{align*}
where $k$ is measured in W/(m\,K). The mean value of the absolute relative error is $0.39\%$ and its maximum is $1.64\%$.

\begin{figure*}[!ht]
         \centering\lineskip=-4pt
         \subfloat{\includegraphics[width=2.92in]{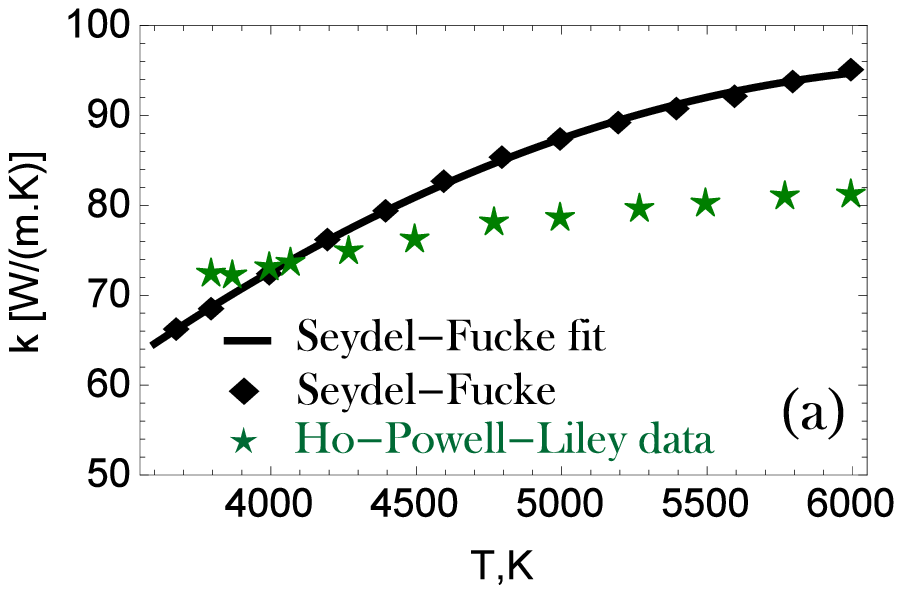}}\qquad
         \subfloat{\includegraphics[width=2.85in]{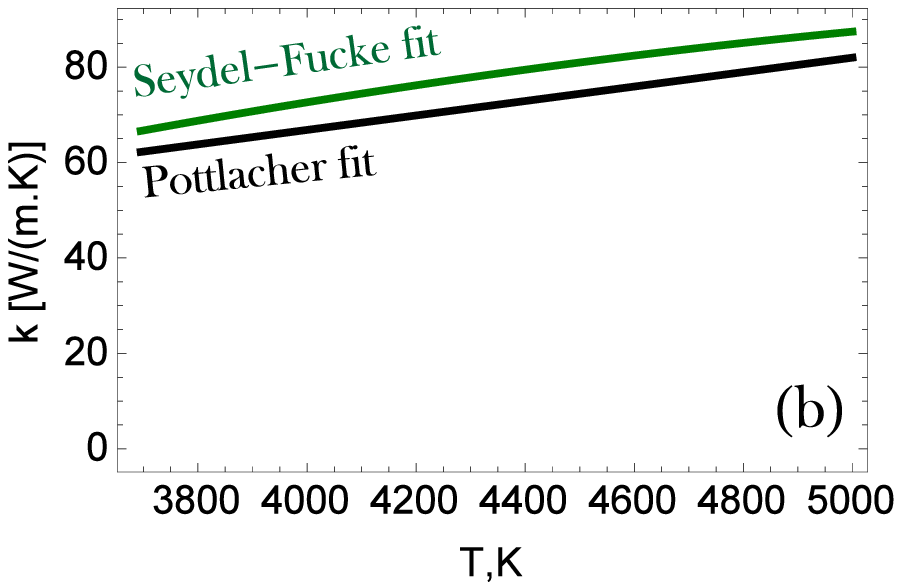}}
\caption{(a) The liquid tungsten thermal conductivity in the range $3695<T(\mathrm{K})<6000$. The data of Seydel--Fucke\,\cite{resistivi04,resistivi05} (together with a quadratic fit) compared with the provisional values provided by Ho--Powell--Liley\,\cite{conductsol1,conductsol2}. (b) The thermal conductivity of liquid tungsten in the temperature range $3695<T(\mathrm{K})<5000$ according to two empirical analytical expressions.}\label{figureWthermalconductivityliquid}
\end{figure*}

\noindent \textbf{Liquid tungsten.} \textbf{(i)} Experimental techniques that directly measure the thermal conductivity are based on formulas that are valid when heat conduction is the only viable mode of heat transfer. Their applicability to high temperature liquid metals such as tungsten is limited due to the increasing importance of convective and radiative heat transfer\,\cite{conductliq1}. \textbf{(ii)} Experimental techniques that measure the thermal diffusivity $\alpha=k/(\rho_{\mathrm{m}}c_{\mathrm{p}})$ can clearly lead to the evaluation of the thermal conductivity\,\cite{conductliq1}. However, post-processing requires the simultaneous knowledge of the mass density and the heat capacity and the measurement uncertainty can be large. \textbf{(iii)} Experimental techniques that measure the electrical resistivity $\rho_{\mathrm{el}}$ can also lead to the evaluation of the thermal conductivity\,\cite{latfusgen01,liquidsbook,conductliq1,conductliq2}. The connecting relation is the Wiedemann-Franz law, $k=L_0T/\rho_{\mathrm{el}}$ with $L_0=2.443\times10^{-8}\,\mathrm{W}\Omega\mathrm{K}^{-2}$ the nominal Lorenz number. In this manner, the abundance of liquid tungsten resistivity data, that have been acquired by dynamic pulse calorimetry, can be translated to thermal conductivity data. The use of the $\rho_{\mathrm{el}}(T)$ fitting expressions with the Wiedemann-Franz law can lead to the propagation of numerical errors. Therefore, when possible, it is preferable that first each resistivity data point is translated to thermal conductivity and that afterwards curve fitting takes place. This procedure has been followed for the Seydel and Fucke measurements\,\cite{resistivi04}. In the original publication, the authors only provide the fitting expression for the resistivity, but their resistivity data have been presented in graphical form in Ref.\cite{resistivi05}. The data have been extracted with the aid of software, they are represented by the average of three different extractions in order to avoid errors due to axis mismatch. The measurements consist of $13$ datapoints from the melting temperature up to $6000\,$K and have been fitted with a quadratic polynomial. The \emph{Seydel--Fucke fit} reads as
\begin{equation*}
k=66.6212+0.02086(T-T_{\mathrm{m}})-3.7585\times10^{-6}(T-T_{\mathrm{m}})^2\,\,\,\,3695\,\mathrm{K}\leq{T}\leq6000\,\mathrm{K}\,,
\end{equation*}
where $k$ is measured in W/(m\,K). The mean value of the absolute relative fitting error is $0.25\%$, see also figure \ref{figureWthermalconductivityliquid}a. Let us compare with the measurements of Pottlacher from melting up to $5000\,$K\,\cite{conductliq5}. The \emph{Pottlacher fit} reads as\,\cite{conductliq5}
\begin{equation*}
k=6.24242+0.01515T\,\,\,\,3695\,\mathrm{K}\leq{T}\leq5000\,\mathrm{K}\,,
\end{equation*}
where $k$ is measured in W/(m\,K). We point out that typical uncertainties in the indirect determination of the thermal conductivity with dynamic pulse calorimetry are $\sim12\%$\,\cite{latfusgen01,conductliq2}. The two fitting functions are plotted in figure \ref{figureWthermalconductivityliquid}b, in their common domain of definition. The deviations are acceptable being $<7\%$. Moreover, we note that the Seydel--Fucke experiments are in better agreement with other recent measurements\,\cite{conductliq6}. Finally, it is worth mentioning that Ho--Powell--Liley provide provisional values for the thermal conductivity of tungsten over its entire liquid range, from the melting up to the critical point\,\cite{conductsol1,conductsol2}. These values were estimated with the phenomenological theory of Grosse, which is based on an empirical hyperbolic relation for the electrical conductivity of liquid metals\,\cite{conductliq7,conductliq8}. As illustrated in figure \ref{figureWthermalconductivityliquid}a and expected due to the oversimplified theoretical analysis, these provisional values are not accurate.

\begin{figure*}[!ht]
         \centering\lineskip=-4pt
         \subfloat{\includegraphics[width=2.90in]{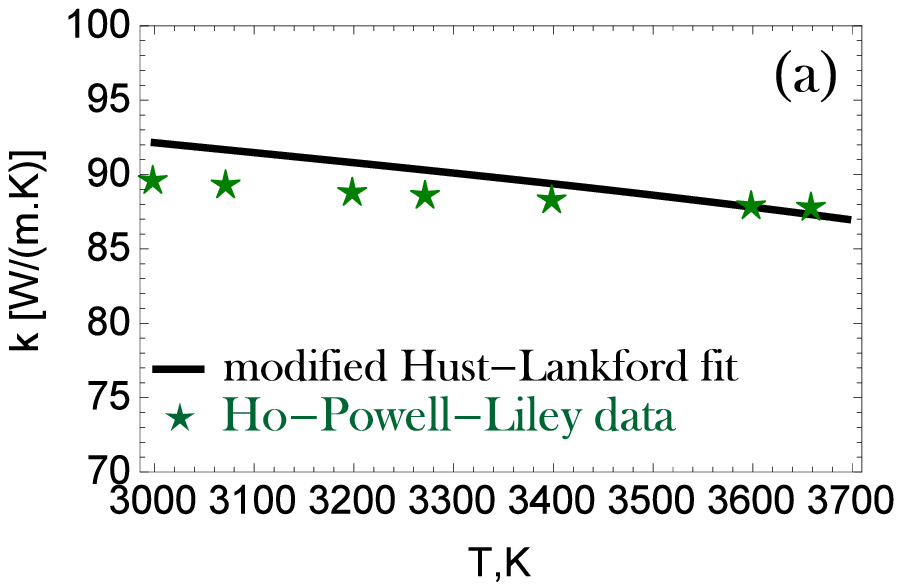}}\qquad
         \subfloat{\includegraphics[width=2.89in]{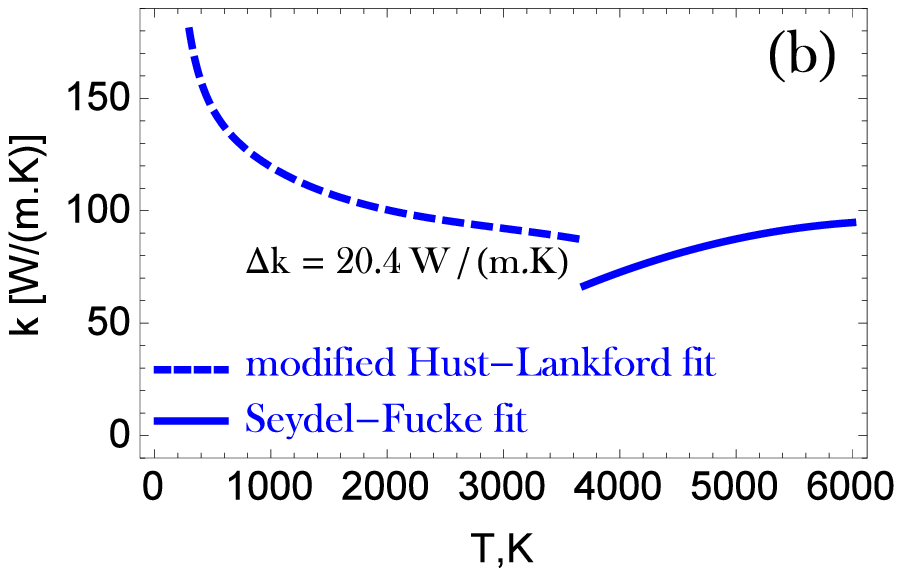}}
\caption{(a) Extrapolation of the modified Hust--Lankford fit in the temperature range $3000<T(\mathrm{K})<3695$ and comparison with the values recommended by Ho--Powell--Liley\,\cite{conductsol1,conductsol2}. (b) The complete recommended analytical description of the W thermal conductivity from $300$ to $6000\,$K.}\label{figureWthermalconductivitytotal}
\end{figure*}

\noindent \textbf{Recommended description.} In order to complete the description, it is necessary to verify that the extrapolation of the modified Hust--Lankford fit in the temperature range $3000<T(\mathrm{K})<3695$ is viable. (i) We have confirmed that the extrapolated values lie very close to the Ho--Powell--Liley recommended dataset\,\cite{conductsol1,conductsol2}, which features seven data-points in this range, see figure \ref{figureWthermalconductivitytotal}a. (ii) We have performed a comparison with the thermal conductivity resulting from the combination of the White--Minges fit for the electrical resistivity\,\cite{resistivi02} with the Wiedemann-Franz law. The agreement was satisfactory. We also note that the two curves overlap when employing $L_{\mathrm{eff}}=1.185L_0$ for the effective Lorenz number. (iii) Overall, the recommended description comprises of the modified Hust--Lankford fit in the temperature range $300<T(\mathrm{K})<3695$ and the Seydel--Fucke fit in the temperature range $3695<T(\mathrm{K})<6000$. See figure \ref{figureWthermalconductivitytotal}b for an illustration. (iv) From the above, we have $k^{\mathrm{s}}\simeq87.0\,$W/(m\,K) and $k^{\mathrm{l}}\simeq66.6\,$W/(m\,K). The resulting discontinuity at the liquid-solid phase transition is $\Delta{k}\simeq20.4\,$W/(m\,K). The large relative magnitude of the discontinuity and the fact that $k^{\mathrm{s}}>k^{\mathrm{l}}$ agrees with results from other refractory metals\,\cite{conductliq1}.

\begin{figure*}[!ht]
         \centering\lineskip=-4pt
         \subfloat{\includegraphics[width=2.83in]{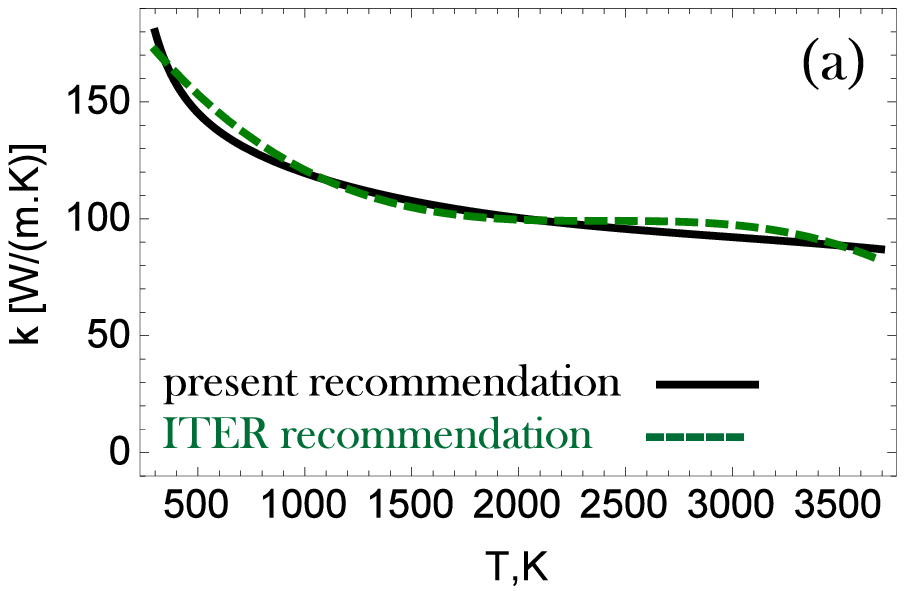}}\qquad
         \subfloat{\includegraphics[width=2.90in]{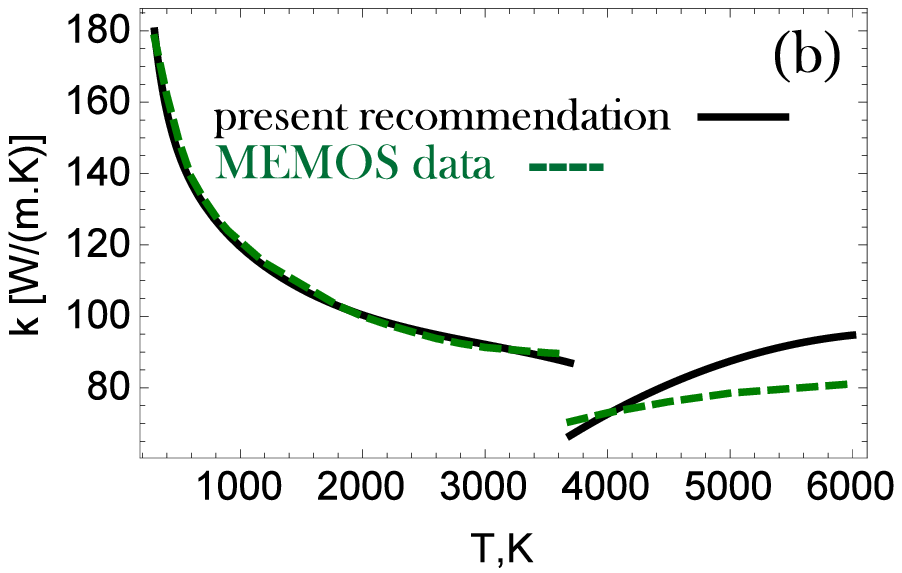}}
\caption{(a) Comparison of the recommended analytical description of the W thermal conductivity with the ITER database recommendation in the temperature range from $300$ to $3695\,$K. (b) Comparison of the recommended analytical description of the W thermal conductivity with the MEMOS code description in the temperature range from $300$ to $6000\,$K.}\label{figureWthermalconductivitycomp}
\end{figure*}

\noindent \textbf{Comparison with the fusion literature}. \emph{In the ITER database}; a cubic polynomial expression is recommended which is valid in the range $273-3653\,$K\,\cite{introduct05}. It originates from fitting to a synthetic dataset, whose high temperature part is heavily based on the recommended dataset provided in the classic 1970 compendium by Touloukian\,\cite{Touloukian1}. As illustrated in figure \ref{figureWthermalconductivitycomp}a, the ITER fit agrees well with our recommended description from the room temperature up to the melting point. However, the ITER fit is characterized by two rather un-physical inflection points, the local sign switching of $\partial{k}/\partial{T}$ might influence thermal modelling due to the $\nabla\cdot(k\nabla{T})$ term in the heat conduction equation. \emph{In the MEMOS code}; Touloukian's recommended dataset is implemented for interpolations in the solid state, whereas the Ho--Powell--Liley provisional dataset is implemented for interpolations in the liquid state\,\cite{introduct08}. As evident from figure \ref{figureWthermalconductivitycomp}b and previous comparisons, the thermal conductivity of solid tungsten is accurately described, while it is mainly underestimated for liquid tungsten. The deviations increase towards the boiling point but never exceed $20\%$.

\subsection{The mass density}\label{density}

\noindent \textbf{Solid tungsten.} The analysis of White and Minges is based on a synthetic dataset constructed from eleven sets of measurements above $300\,$K and three sets of measurements below $300\,$K\,\cite{resistivi02}. They provide a least-squares polynomial fit for the linear expansion coefficient $\alpha_{\mathrm{l}}=(1/l_0)(dl/dT)$, where $l_0$ is the length measured at the room temperature ($T_0=293.15\,$K), that is valid from $300\,$K up to $3500\,$K. The linear expansion coefficient fit reads as
\begin{align*}
\alpha_{\mathrm{l}}=3.873+2.562\times10^{-3}T-2.8613\times10^{-6}T^2+1.9862\times10^{-9}T^3-0.58608\times10^{-12}T^4+0.070586\times10^{-15}T^5\,,
\end{align*}
where $\alpha_{\mathrm{l}}$ is measured in $10^{-6}\,\mathrm{K}^{-1}$. These authors only provide tabulated data for the relative change in the linear dimension $\Delta{l}/l_0=(l-l_0)/l_0$. An analytical expression for the normalized linear dimension can either be calculated from the relation $l/l_0=1+\int_{T_0}^{T}\alpha_{\mathrm{l}}(T^{\prime})dT^{\prime}$ or by least-squares fitting the tabulated data and employing $l/l_0=1+\Delta{l}/l_0$. The normalized linear dimension fit reads as
\begin{align*}
\frac{l}{l_0}=1+4.64942\times10^{-6}(T-T_0)+2.99884\times10^{-11}(T-T_0)^2+1.95525\times10^{-13}(T-T_0)^3\,.
\end{align*}
In the case of isotropic thermal expansion for a cubic metal such as tungsten, we have $V/V_0=(l/l_0)^3$ for the volume expansion. The specific volume fit reads as
\begin{align*}
\frac{V}{V_0}=1+1.4016\times10^{-5}(T-T_0)+4.4004\times10^{-11}(T-T_0)^2+6.3724\times10^{-13}(T-T_0)^3\,.
\end{align*}
Finally, the dependence of the mass density of solid tungsten on the temperature can be evaluated by employing $\rho_{\mathrm{m0}}=19.25\,$g\,cm$^{-3}$ for the room temperature mass density and $V/V_0=\rho_{\mathrm{m0}}/\rho_{\mathrm{m}}$ as imposed by mass conservation. The \emph{White--Minges fit} reads as
\begin{align*}
\rho_{\mathrm{m}}=19.25-2.66207\times10^{-4}(T-T_0)-3.0595\times10^{-9}(T-T_0)^2-9.5185\times10^{-12}(T-T_0)^3\,\,\,\,300\leq{T}(\mathrm{K})\leq3400\,,
\end{align*}
where $\rho_{\mathrm{m}}$ is measured in g\,cm$^{-3}$.

\begin{figure*}[!ht]
         \centering\lineskip=-4pt
         \subfloat{\includegraphics[width=2.75in]{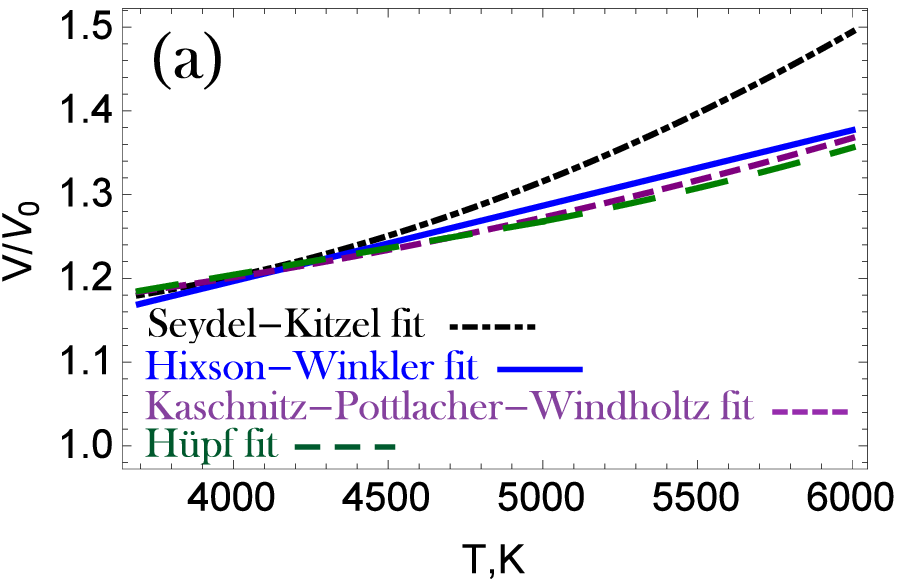}}\qquad
         \subfloat{\includegraphics[width=2.85in]{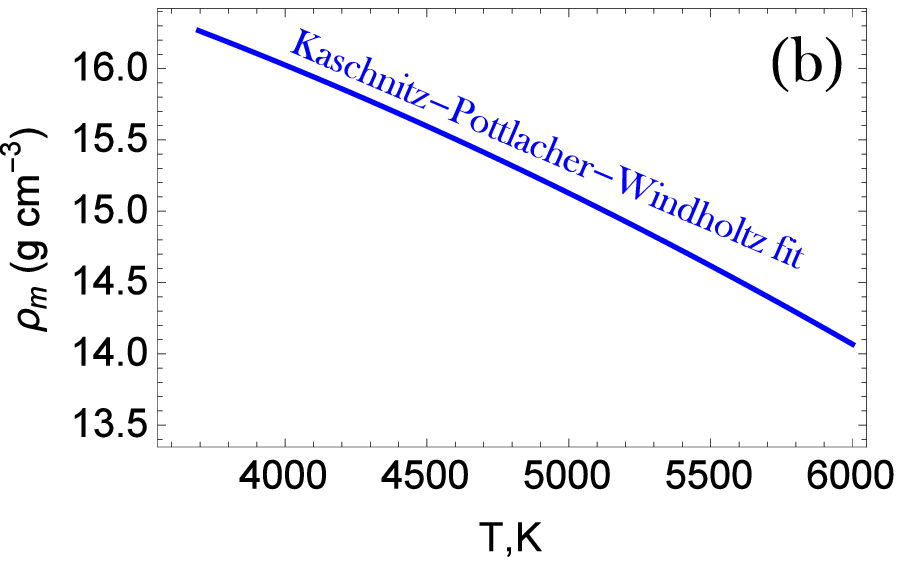}}
\caption{(a) The liquid tungsten thermal volume expansion in the range $3695<T(\mathrm{K})<6000$ according to four empirical analytical expressions\,\cite{latfusexp09,latfusexp10,resistivi06,volumeliqu1}. (b) The mass density of liquid tungsten in the range $3695<T(\mathrm{K})<6000$ according to the Kaschnitz--Pottlacher--Windholz fit\,\cite{latfusexp10}.}\label{figureWthermalexpansionliquid}
\end{figure*}

\noindent \textbf{Liquid tungsten.} So far we have employed the Seydel and Fucke measurements\,\cite{resistivi04} for the electrical resistivity and the thermal conductivity of liquid tungsten. It would be consistent to employ the volume expansion data originating from the same experimental group, provided of course that they are reliable. \textbf{(i)} Seydel and Kitzel have provided thermal volume expansion data for five refractory metals (Ti, V, Mo, Pd, W) from their melting up to their boiling point\,\cite{volumeliqu1}. They have successfully fitted the specific volume of tungsten to a quadratic polynomial. The \emph{Seydel--Kitzel fit} reads as
\begin{equation*}
\frac{V}{V_0}=1.18+6.20\times10^{-5}(T-T_{\mathrm{m}})+3.23\times10^{-8}(T-T_{\mathrm{m}})^2\,,
\end{equation*}
where $V_0$ is the tungsten specific volume in room temperature. It is worth noting that the Seydel--Kitzel fit has been singled out as the recommended expression in specialized reviews\,\cite{volumeliqu2}. \textbf{(ii)} Hixson and Winkler have measured the specific volume of liquid tungsten in the range $3695\leq{T}(\mathrm{K})\leq5700$\,\cite{latfusexp09}. They have provided linear expressions for the specific volume as a function of the enthalpy and for the enthalpy as a function of the temperature. Combining their expressions, we acquire the \emph{Hixson--Winkler fit} that reads as
\begin{equation*}
\frac{V}{V_0}=0.83634+0.901\times10^{-4}T\,,
\end{equation*}
where $V_0$ is the tungsten specific volume in room temperature. \textbf{(iii)} Kaschnitz, Pottlacher and Windholz have carried out similar measurements without providing fitting expressions\,\cite{latfusexp10}. However, the analytical fit of the specific volume as a function of the temperature has been plotted in a figure. We digitized this figure in the temperature range $3695\leq{T}(\mathrm{K})\leq6000$ with steps of $100\,$K and we least-square fitted the resulting dataset to a quadratic polynomial. The \emph{Kaschnitz--Pottlacher--Windholz fit} reads as
\begin{equation*}
\frac{V}{V_0}=1.184+5.27\times10^{-5}(T-T_{\mathrm{m}})+1.17\times10^{-8}(T-T_{\mathrm{m}})^2\,,
\end{equation*}
where again $V_0$ is the tungsten specific volume in room temperature. The mean value of the absolute relative fitting error is $0.05\%$. \textbf{(iv)} H\"upf \emph{et al.} have also measured the volume expansion of five refractory liquid metals (V, Nb, Ta, Mo, W)\,\cite{resistivi06}. We note that the authors provided a fit for the quantity $D^2/D_0^2$ as a function of the temperature, where $D$ denotes the wire diameter. Under rapid heating the melted wire expands solely in the radial direction, which implies that its volume is proportional to the cross-section and thus $V/V_0=D^2/D_0^2$\,\cite{conductliq2,volumeliqu3}. The fitting expression consists of two polynomial branches, but it is continuous at the branch point. The \emph{H\"upf fit} reads as
\begin{equation*}
\frac{V}{V_0}=
\begin{cases}
0.95062+6.344\times10^{-5}T\,\,\,\,3695\leq{T}(\mathrm{K})\leq5000,\\
1.34989-1.0333\times10^{-4}T+1.73957\times10^{-8}T^2\,\,\,\,5000\leq{T}(\mathrm{K})\leq6000,
\end{cases}
\end{equation*}
where again $V_0$ is the tungsten specific volume in room temperature. The four fits are illustrated in figure \ref{figureWthermalexpansionliquid}a. It is evident that the Seydel--Kitzel fit greatly overestimates the volume expansion for very high temperatures with the deviations from the other curves starting from $4500\,$K. The cause of this overestimation was investigated in a seminal paper by Ivanov, Lebedev and Savvatimskii\,\cite{volumeliqu3}; All the aforementioned experiments were based on the resistive pulse heating technique and the volume expansion measurements were carried out by recording the temporal evolution of the shadow the sample produced after illumination with a radiation source either in a dense gas or in a liquid. Only Seydel and Kitzel performed their experiments in water\,\cite{volumeliqu1}. In that case, a layer of vapour surrounded the sample with its thickness determined by the sample temperature and its rapid evolution. Since vapor possesses a refractive index smaller than that of water, the vapor layer caused the shadow image to expand and was responsible for the overestimation. The correctness of the other fits was confirmed by the same authors by measurements of the thermal expansion of liquid tungsten with two alternative independent techniques, the capillary method and the probe method\,\cite{volumeliqu1}. From figure \ref{figureWthermalexpansionliquid}a, it is also evident that, close to the melting point, the H\"upf fit deviates from the other curves. Combining the above and considering the more limited temperature range of the Hixson--Winkler fit, we conclude that the Kaschnitz--Pottlacher--Windholz fit is the most appropriate. It is preferable to convert this fit to an analytical expression for the mass density. Using $\rho_{\mathrm{m0}}=19.25\,$g\,cm$^{-3}$ for the room temperature mass density of tungsten and $V/V_0=\rho_{\mathrm{m0}}/\rho_{\mathrm{m}}$, we acquire the \emph{Kaschnitz--Pottlacher--Windholz fit} for the mass density
\begin{equation*}
\rho_{\mathrm{m}}=16.267-7.679\times10^{-4}(T-T_{\mathrm{m}})-8.091\times10^{-8}(T-T_{\mathrm{m}})^2\,\,\,\,3695\leq{T}(\mathrm{K})\leq6000\,,
\end{equation*}
where $\rho_{\mathrm{m}}$ is measured in g\,cm$^{-3}$. This fit is illustrated in figure \ref{figureWthermalexpansionliquid}. The density of liquid tungsten at the melting point is $\rho_{\mathrm{m}}^{\mathrm{l}}=16.267\,$g\,cm$^{-3}$, which is very close to typical values recommended in handbooks.

\begin{figure*}[!ht]
         \centering\lineskip=-4pt
         \subfloat{\includegraphics[width=2.85in]{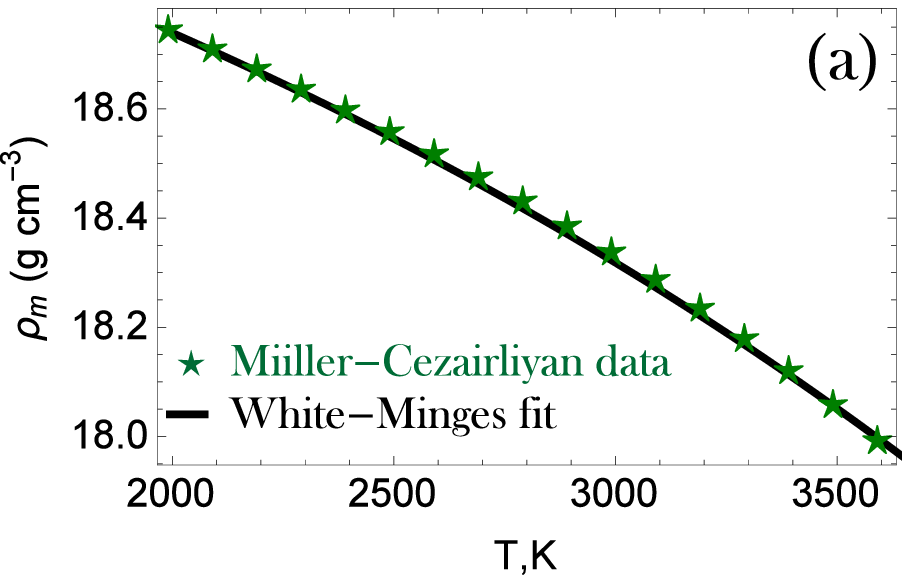}}\qquad
         \subfloat{\includegraphics[width=2.85in]{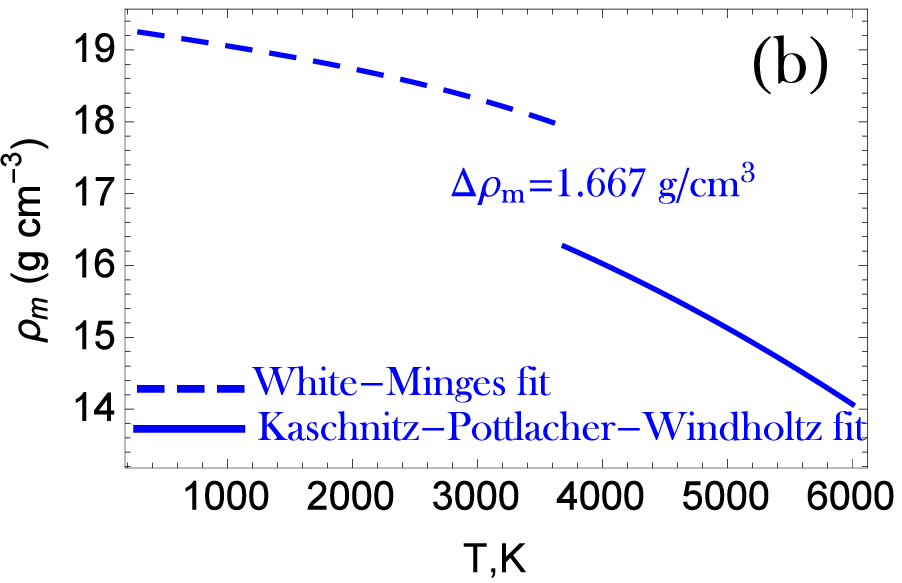}}
\caption{(a) Comparison of the White--Minges fit close to the tungsten melting point, $2000<T(\mathrm{K})<3695$, with the dedicated high temperature measurements of Miiller and Cezairliyan\,\cite{volumeliqu4}. (b) The complete recommended analytical description of the tungsten mass density from $300$ to $6000\,$K.}\label{figureWthermalexpansion}
\end{figure*}

\noindent \textbf{Recommended description.} In order to complete the description, it is necessary to verify that the White--Minges fit is reliable at high temperatures close to the melting point. Miiller and Cezairliyan had employed a precise high-speed interferometric technique for the measurement of the thermal expansion of tungsten from $1500\,$K up to the melting point\,\cite{volumeliqu4}. The maximum uncertainty in the measured linear expansion was estimated to be $\sim1\%$ at $2000\,$K and $\sim2\%$ at $3600\,$K. From figure \ref{figureWthermalexpansion}a, it is clear that their experimental results are nearly indistinguishable from the White--Minges fit. Overall, the recommended description comprises of the White--Minges fit in the temperature range $300<T(\mathrm{K})<3695$ and the Kaschnitz--Pottlacher--Windholz fit in the temperature range $3695<T(\mathrm{K})<6000$. See figure \ref{figureWthermalexpansion}b for an illustration. From the above, we have $\rho_{\mathrm{m}}^{\mathrm{s}}=17.934\,$g\,cm$^{-3}$ and $\rho_{\mathrm{m}}^{\mathrm{l}}=16.267\,$g\,cm$^{-3}$. The resulting discontinuity at the liquid-solid phase transition is $\Delta\rho_{\mathrm{m}}=1.667\,$g\,cm$^{-3}$. As expected we have $\rho_{\mathrm{m}}^{\mathrm{s}}>\rho_{\mathrm{m}}^{\mathrm{l}}$ similar to most metals\,\cite{liquidsbook}. It is worth noting that the large relative magnitude of the discontinuity implies a rather large volume expansion during melting compared to other bcc metals\,\cite{liquidsbook}.

\subsection{The surface tension}\label{tension}

\noindent \textbf{Significance.} The surface tension is a fundamental physical quantity in various plasma-material interaction phenomena that are important for fusion devices: \emph{(i) Droplet generation.} The velocity difference at the interface between the edge plasma and the melt layer leads to the development of the Kelvin-Helmholtz instability and the growth of surface waves whose subsequent breakup can result to metallic droplet ejection into the plasma\,\cite{surfacegen1,surfacegen2}. The surface tension impedes the growth of the K-H instability by providing the restoring force that stabilizes short wavelength perturbations\,\cite{surfacegen3}. \emph{(ii) Droplet disintegration}. The shape of charged spherical droplets is subject to distortions due to electrostatic pressure\,\cite{surfacegen4}. The surface tension counteracts the electrostatic pressure which tends to rip the droplets apart. The application of the classical Rayleigh linear analysis for metallic droplets embedded in fusion plasmas leads to a threshold radius below which electrostatic disruption occurs and whose value is inversely proportional to the surface tension\,\cite{surfacegen5}. \emph{(iii) Melt-layer motion}. Surface tension gradients stemming from surface temperature gradients naturally result to thermo-capillary flows that can influence macroscopic melt-layer motion. Since surface tension enters the mathematical description through the boundary condition that expresses the balance between the tangential hydrodynamic stress and the surface tension gradient, its effect is more transparent when inspecting the Navier-Stokes system within the shallow water approximation, where it contributes a source term proportional to $\left(\partial\sigma/\partial{T}\right)\left(\nabla{T}\right)$ to the non-normal liquid metal velocity components\,\cite{introduct08}.

\noindent \textbf{Liquid metals.} Conventional techniques can be utilized for the measurement of the surface tension of liquid metals such as the maximum bubble pressure method, the sessile drop method and the pendant drop - drop weight method\,\cite{surfaceten1}. For melts of refractory metals, container-less (or non-contact) methods and particularly levitating drop methods are required in order to eliminate the possibility of chemical reactions between the melt and crucibles or substrates\,\cite{surfaceten2,surfaceten3,surfaceten4}. Different variants of the levitating drop method have been developed such as aerodynamic, optical, electrostatic and electromagnetic levitation\,\cite{surfaceten3,surfaceten4}. The experimental results originating from electrostatic levitation measurements are generally considered to be more accurate\,\cite{surfaceten2} due to the inherent advantages of this method\,\cite{surfaceten4,surfaceten5}. The electrostatic levitation method is based on lifting a small charged material sample with the aid of electrostatic fields, melting the sample with the aid of lasers, inducing shape oscillations by applying a small amplitude ac modulation to the field, recording the oscillating frequency as well as the amplitude damping of the drop shape profile, which provide the surface tension and the viscosity\,\cite{surfaceten5}. It is worth noting that the temperature dependence of the liquid metal surface tension has also been extensively studied because of the aforementioned thermo-capillary Marangoni flows. In general, it is assumed that the dependence of the surface tension of pure liquid metals on the temperature is linear\,\cite{surfaceten1,surfaceten2,surfaceten4,liquidbook2}. This linearity is not imposed by generic theoretical arguments, but more likely stems from the limited temperature range of the experiments and the insufficient accuracy of the measurements. The basic constraint imposed by thermodynamics is that the surface tension reduces to zero at the critical point\,\cite{liquidsbook}. These remarks imply that the temperature coefficient is always negative; positive values have been measured but - most of the times - they can be attributed to impurity effects or non-equilibrium conditions\,\cite{liquidsbook}.

\noindent \textbf{Liquid tungsten.} Numerous reviews dedicated to experimental measurements of the surface tension of liquid metals can be encountered in the literature\,\cite{surfaceten1,surfaceten2,surfaceten6,surfaceten7}. In these compilations, very few data can be found for the surface tension of tungsten at the melting point and no measurements can be found for the temperature dependence of the tungsten surface tension. Fortunately, very recent experiments were carried out by Paradis \emph{et al.} with the electrostatic levitation method\,\cite{surfaceten8}. The surface tension was measured for liquid tungsten barely above the melting point and in the under-cooled phase, $3360<T(\mathrm{K})<3700$. The temperature interval of $350\,$K can be considered as adequate for the determination of the temperature coefficient. A linear fit of the form $\sigma=\sigma_{\mathrm{m}}-\beta(T-T_{\mathrm{m}})$ provided an accurate description of the data, which - in absence of other measurements - needs to be extrapolated in the entire liquid phase. The \emph{Paradis fit} reads as\,\cite{surfaceten8}
\begin{equation*}
\sigma=2.48-0.31\times10^{-3}(T-T_{\mathrm{m}})
\end{equation*}
where $\sigma$ is measured in N/m. The uncertainties in the least square fit coefficients are $\sim10\%\,(\sigma_{\mathrm{m}})$ and $\sim25\%\,(\beta)$. The surface tension at the melting point $\sigma_{\mathrm{m}}$ displays a strong agreement with previous measurements, as seen in Table \ref{tablesurfacetensionmelting}. We shall check how physical is the experimental value of the linear coefficient $\beta$ by extrapolating at very high temperatures and determining the critical point temperature from $\sigma=0$. The result is $T_{\mathrm{c}}\simeq11700\,$K. There is a remarkable agreement with numerous  estimates of the tungsten critical point. In particular; the Guldberg rule leads to $12277\,$K, the Likalter equation of state leads to $12466\,$K, the Goldstein scaling leads to $11852\,$K and dynamic experiments using exploding wires lead on average to $12195\,$K\,\cite{criticalpoi}.

\begin{table*}[!t]
  \centering
  \caption{The surface tension of tungsten at the melting temperature according to dedicated experiments. The dataset of Allen\,\cite{surfaceexp2} has been corrected for the liquid mass density following Ref.\cite{surfaceexp4}, since the exact experimental output in the pendant drop - drop weight method is the ratio $\sigma/\rho_{\mathrm{m}}$ and the room-temperature tungsten density was employed in the original work.}\label{tablesurfacetensionmelting}
  \begin{tabular}{c c c c c} \hline
Investigators                & \,\,Reference\,\,  & \,\,Year\,\, & \,\,$\sigma_{\mathrm{m}}$ (N/m)\,\, & \,\,Experimental method\,\,                    \\ \hline
Caverly                      & \cite{surfaceexp1} &    1957      & 2.300                               & \,\,pendant drop - drop weight\,\,             \\
Allen                        & \cite{surfaceexp2} &    1963      & 2.355                               & \,\,pendant drop - drop weight\,\,             \\
Martsenyuk \emph{et al.}\,\, & \cite{surfaceexp3} &    1974      & 2.316                               & \,\,pendant drop - drop weight\,\,             \\
Vinet \emph{et al.}          & \cite{surfaceexp4} &    1993      & 2.310                               & \,\,pendant drop - drop weight\,\,             \\
Paradis \emph{et al.}        & \cite{surfaceten8} &    2005      & 2.480                               & \,\,electrostatic levitation\,\,               \\ \hline \hline
\end{tabular}
\end{table*}

\subsection{The dynamic viscosity}\label{viscosity}

\noindent \textbf{Liquid metals.} Conventional experimental techniques can be employed for the measurement of the dynamic viscosity of liquid metals such as the capillary method, the oscillating vessel method, the rotating cylinder method\,\cite{liquidsbook,viscosityg1,viscosityg2}. For melts of refractory metals, non-contact techniques such as the electrostatic levitation method are preferred due to the high melting temperatures and the enhanced reactivity at elevated temperatures\,\cite{viscosityg3,viscosityg4}. In general, it is assumed that the dependence of the dynamic viscosity of pure liquid metals on the temperature is of the Arrhenius form\,\cite{liquidsbook,liquidbook2}, \emph{i.e.} $\mu(T)=\mu_0\exp{\left[E_{\mathrm{a}}/(RT)\right]}$ with $E_{\mathrm{a}}$ the activation energy for viscous flow, $\mu_0$ the pre-exponential viscosity and $R=8.314\,$J/(mol$\cdot$K) the ideal gas constant. It should also be emphasized that, within some limitations, the dynamic viscosity and the surface tension are connected by a rigorous statistical mechanics relation. The Fowler formula for the surface tension of liquids reads as $\sigma(T)=(\pi{n}^2/8)\int_0^{\infty}r^4g(r,T)[d\phi(r,T)/dr]dr$, where $g(r,T)$ is the pair correlation function, $\phi(r,T)$ is the effective pair interaction potential and $n$ is the particle number density\,\cite{viscosityg5,viscosityg6}. The Born-Green formula for the viscosity of liquids reads as $\mu(T)=\sqrt{m/(k_{\mathrm{B}}T)}(2\pi{n}^2/15)\int_0^{\infty}r^4g(r,T)[d\phi(r,T)/dr]dr$, with $m$ the particle atomic mass\,\cite{viscosityg7}. Dividing by parts, the Fowler-Born-Green formula emerges, $\mu(T)=(16/15)\sqrt{m/(k_{\mathrm{B}}T)}\sigma(T)$\,\cite{viscosityg8,viscosityg9,viscosityg0}. The fundamental assumptions behind the Fowler formula and the Born-Green formula determine the applicability range of this simple elegant formula\,\cite{viscosityg0}, which has proved to be very accurate for elemental liquid metals\,\cite{liquidsbook}.

\begin{figure*}[!ht]
         \centering\lineskip=-4pt
         \subfloat{\includegraphics[width=2.90in]{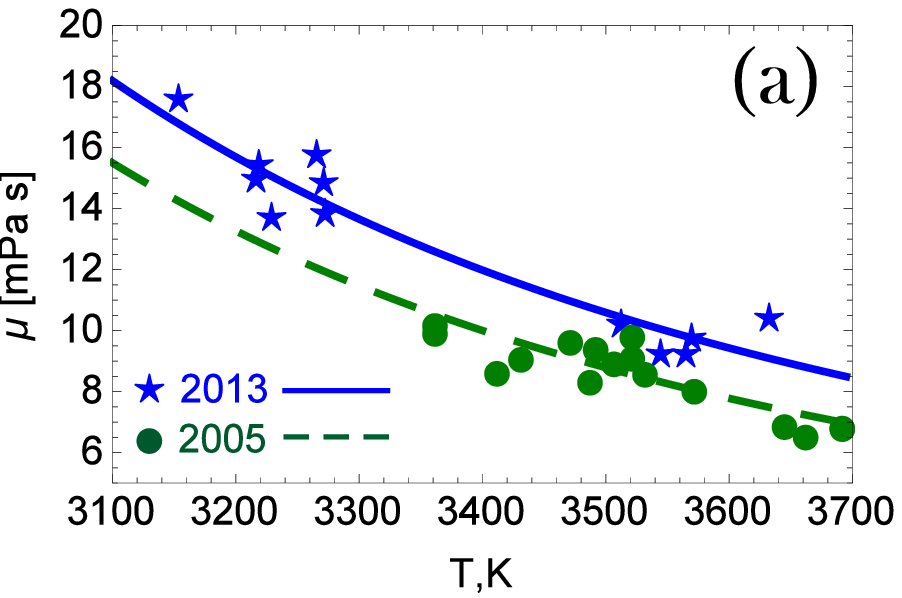}}\qquad
         \subfloat{\includegraphics[width=3.17in]{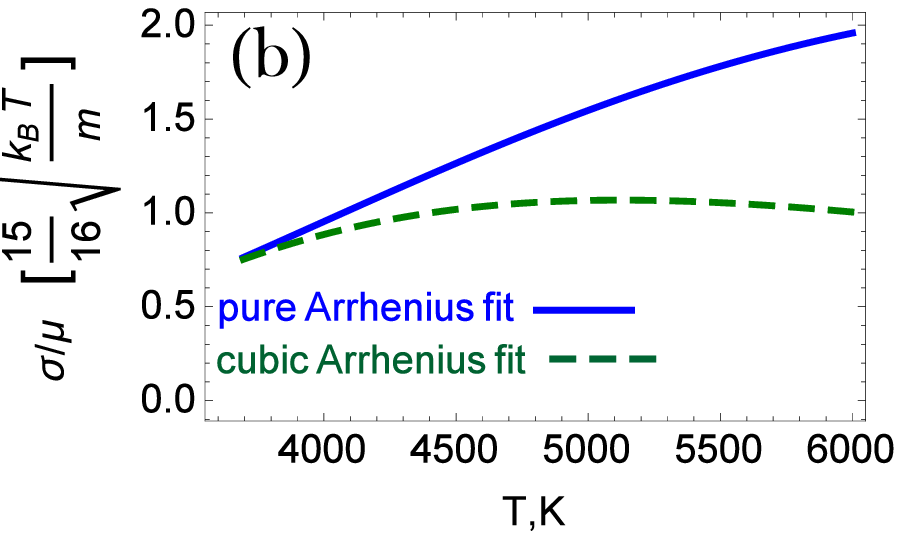}}
\caption{ (a) The viscosity of liquid tungsten in the under-cooled phase; experimental results and Arrhenius least square fits\,\cite{viscositye4,viscositye5}. (b) The validity of the Fowler-Born-Green formula across the liquid phase of tungsten adopting the linear expression for the surface tension and either a pure Arrhenius or a cubic Arrhenius expression for the viscosity.}\label{figureWdynamicviscosity}
\end{figure*}

\noindent \textbf{Liquid tungsten.} Numerous works that have reviewed experimental data for the viscosity of liquid metals, elemental but also alloys, can be encountered in the literature\,\cite{latfusgen02,liquidbook2,viscositye1,viscositye2,viscositye3}. Similar to the case of surface tension, in these compilations, very few data can be found for the viscosity of tungsten at the melting point and nearly no measurements for its temperature dependence. The only exception are very recent experiments that were carried out by Ishikawa \emph{et al.} with the electrostatic levitation method\,\cite{viscositye4,viscositye5}. The measurements reported in Ref.\cite{viscositye5} will be considered in greater detail, since it has been concluded that the measurements of Ref.\cite{viscositye4} were affected by sample positioning forces. In Ref.\cite{viscositye5}, the viscosity was measured for liquid tungsten in the under-cooled phase, $3155<T(\mathrm{K})<3634$. The temperature interval of $480\,$K can be considered as adequate for the determination of the temperature dependence. An Arrhenius fit of the form $\mu=\mu_0\exp{\left[E_{\mathrm{a}}/(RT)\right]}$ provided an accurate description of the data, which - in absence of other measurements - needs to be extrapolated in the entire liquid phase. The \emph{Ishikawa fit} reads as\,\cite{viscositye5}
\begin{align*}
\mu=0.16\times10^{-3}\exp{\left(3.9713\frac{T_{\mathrm{m}}}{T}\right)}\,,
\end{align*}
where $\mu$ is measured in Pa\,s. This expression corresponds to an activation energy $E_{\mathrm{a}}=122\times10^{3}\,$J/mol which has been determined by least square fitting with a $20\%$ uncertainty. The extrapolated value of the viscosity at the melting point is $\mu(T_{\mathrm{m}})=8.5\times10^{-3}\,$Pa\,s close to the experimental value $\mu(T_{\mathrm{m}})=7.0\times10^{-3}\,$Pa\,s provided in the literature\,\cite{liquidbook2}. Both measurements of Ishikawa et al.\,\cite{viscositye4,viscositye5} are illustrated in figure \ref{figureWdynamicviscosity}a together with the least square fitted Arrhenius expressions. In the absence of other experimental results, the Fowler-Born-Green formula provides the only way to cross-check the adopted measurements. In figure \ref{figureWdynamicviscosity}b, the ratio of $\sigma(T)/\mu(T)$, where $\sigma(T)$ follows the Paradis fit and $\mu(T)$ follows the Ishikawa fit, is expressed in units of $(15/16)\sqrt{(k_{\mathrm{B}}T)/m}$ and plotted as a function of the temperature. The quantity does not diverge strongly from unity, especially taking into account the experimental uncertainties and the wide extrapolations in the viscosity as well as the surface tension. It is worth investigating whether fitting expressions other than the pure Arrhenius form can equally fit the experimental data, but provide a better agreement with the Fowler-Born-Green formula. A cubic Arrhenius fit,
\begin{align*}
\mu=2.76\times10^{-3}\exp{\left[1.1362\left(\frac{T_{\mathrm{m}}}{T}\right)^3\right]}\,,
\end{align*}
where $\mu$ is measured in Pa\,s, fulfills these two criteria. It is impossible to determine whether the extrapolated Arrhenius fit for the viscosity or the extrapolated linear fit for the surface tension are responsible for the deviations from the Fowler-Born-Green formula, not to mention that this formula should not be exactly obeyed across the liquid phase. Therefore, we still recommend the use of the pure Arrhenius fit. The aim of this comparison was to highlight the need for tungsten surface tension and viscosity measurements in larger temperature ranges and for temperatures exceeding the melting point.

\section{Discussion and summary}

\subsection{Complications in burning fusion plasma environments}\label{complications}

\noindent The recommended analytical expressions are nearly exclusively based on experimental results for pure polycrystalline tungsten. Nevertheless, unless rather extreme cases are considered, microstructural details and impurity concentrations should have a negligible influence on the thermophysical properties of interest. Even in the case of pure surface quantities that are very sensitive to adsorbates, such as the surface tension in the liquid phase, the volatility of low-Z contaminants at elevated temperatures guarantees a limited effect. Considering the hostile edge plasma environment of magnetic fusion reactors, it is inevitable that complications arise which should be discussed in further detail. These mainly concern the possible impact of external magnetic fields, plasma contamination, beryllium-tungsten alloying and neutron irradiation.

(a) \emph{Magnetic field effects.} The prominent role of the de-localized valence electrons in charge and heat transport implies that strong external magnetic fields could influence the magnitude and alter the isotropic nature of thermophysical properties such as the thermal conductivity or the electrical resistivity. However, even for high field strengths, magnetic field effects can be expected to be very weak for tungsten, since the mean free paths are much smaller than the Larmor radii due to the enormous density of the scattering centers. Order of magnitude estimates can be performed with the aid of the elementary Drude model, \emph{i.e.} a single particle description with friction described by the relaxation time approximation\,\cite{conductsol4}. The valence electron density is $n_{\mathrm{e}}\simeq1.3\times10^{29}\,$m$^{-3}$ and the mean time between collisions is given by $\tau_{\mathrm{e}}=m_{\mathrm{e}}/(n_{\mathrm{e}}e^2\rho_{\mathrm{el}})$, which lead to $\omega_{\mathrm{ce}}\tau_{\mathrm{e}}=B/(en_{\mathrm{e}}\rho_{\mathrm{el}})$, where $\omega_{\mathrm{ce}}$ denotes the cyclotron frequency of the valence electrons. For $B=6\,$T and room temperature, we have $\omega_{\mathrm{ce}}\tau_{\mathrm{e}}\sim5\times10^{-3}$. Within the Drude model, the relative electrical resistivity increase is $\Delta\rho_{\mathrm{el}}/\rho_{\mathrm{el}}=(\omega_{\mathrm{ce}}\tau_{\mathrm{e}})^2$\,\cite{conductsol4} which is clearly negligible.

(b) \emph{Plasma contaminants.} As analyzed in the previous section, for high impurity or imperfection concentrations, interaction with defects may dominate the valence electron transport and thus drastically modify quantities such as the thermal conductivity or the electrical resistivity. High hydrogen and helium atom concentrations are unavoidable in the surface proximity of plasma-facing components, owing to the implantation and trapping of the incident plasma ions. It has been documented in the literature that helium bubble and tungsten fuzz formation lead to the degradation of the local thermal properties\,\cite{outroplasm1,outroplasm2}, as expected from the porous or fiber-like surface morphology. In particular, recent thermal conductivity measurements for tungsten damaged by high flux - low energy helium plasma revealed a $80\%$ reduction\,\cite{outroplasm3}. Such phenomena may have an impact on the PFC power-handling capabilities and more systematic measurements need to be carried out in order to document the extent of the thermally degraded near-surface region. However, they are most likely not relevant for repetitive transient melting events. At elevated temperatures (still well below the melting temperature), trapped gas desorption accompanied by nano-structure annealing\,\cite{outroplasm2,outroplasm4} can be expected to strongly limit the effect of plasma contamination from the beginning of the ELM cycle.

(c) \emph{Be-W alloying effects.} Beryllium erosion from the first wall and its transport to the divertor has been well-understood and documented in JET\,\cite{outroplasm5,outroplasmN}. Moreover, the ITER divertor surface is expected to be covered by a thin beryllium layer\,\cite{outroplasm2}. Under the appropriate plasma conditions (so that significant concentrations of beryllium remain locally deposited) and surface temperatures (so that element inter-diffusion is significant) beryllium-tungsten alloys can form, for instance Be$_2$W with $T_{\mathrm{m}}\sim2520\,$K or Be$_{12}$W with $T_{\mathrm{m}}\sim1780\,$K\,\cite{outroplasm2,outroplasm6,outroplasm7,outroplasm8}. As exhibited by the much lower melting points, mixed beryllium-tungsten materials are characterized by thermophysical properties that strongly depend on the alloy stoichiometry. Optimized conditions for the growth of Be-W alloys might occur near the strike point\,\cite{outroplasm9}. Further R\&D is necessary to quantify the local extent of Be-W alloy formation. It is evident though that, unless the thickness of the alloy layer is significant, its presence should not be important for thermal or hydrodynamic modelling in spite of the totally different thermophysical properties compared to tungsten.

\begin{figure*}[!t]
         \centering\lineskip=-4pt
         \subfloat{\includegraphics[width=3.30in]{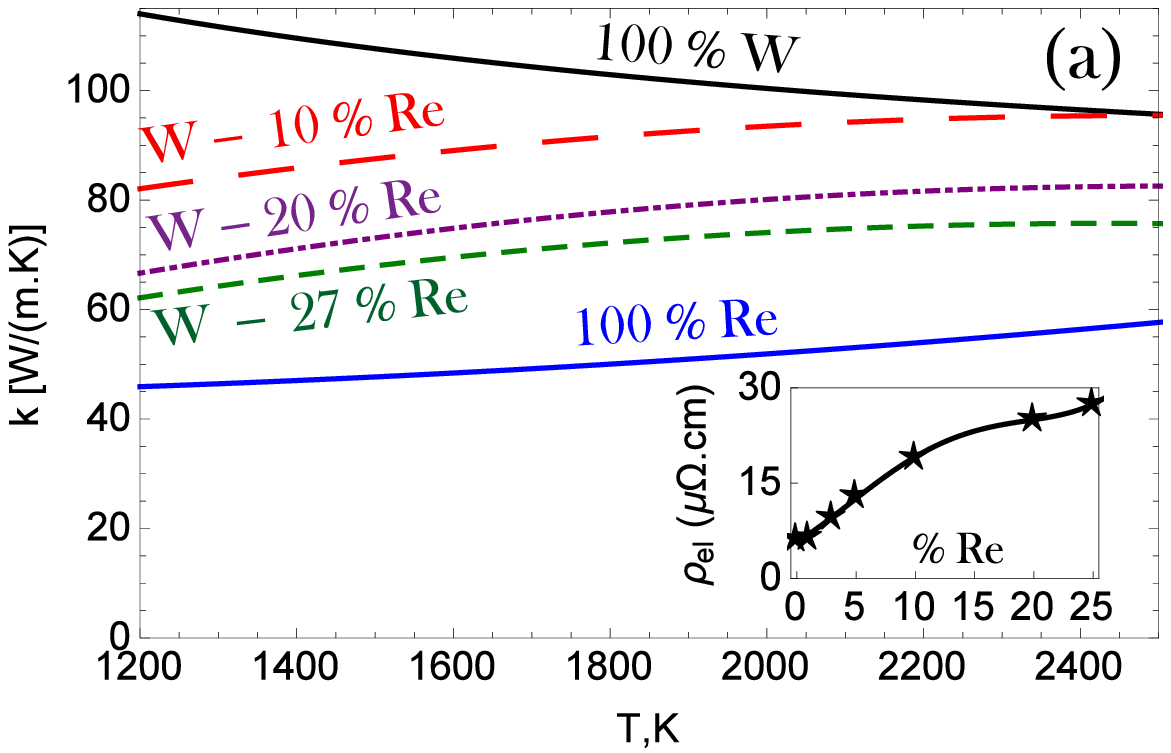}}\qquad
         \subfloat{\includegraphics[width=3.30in]{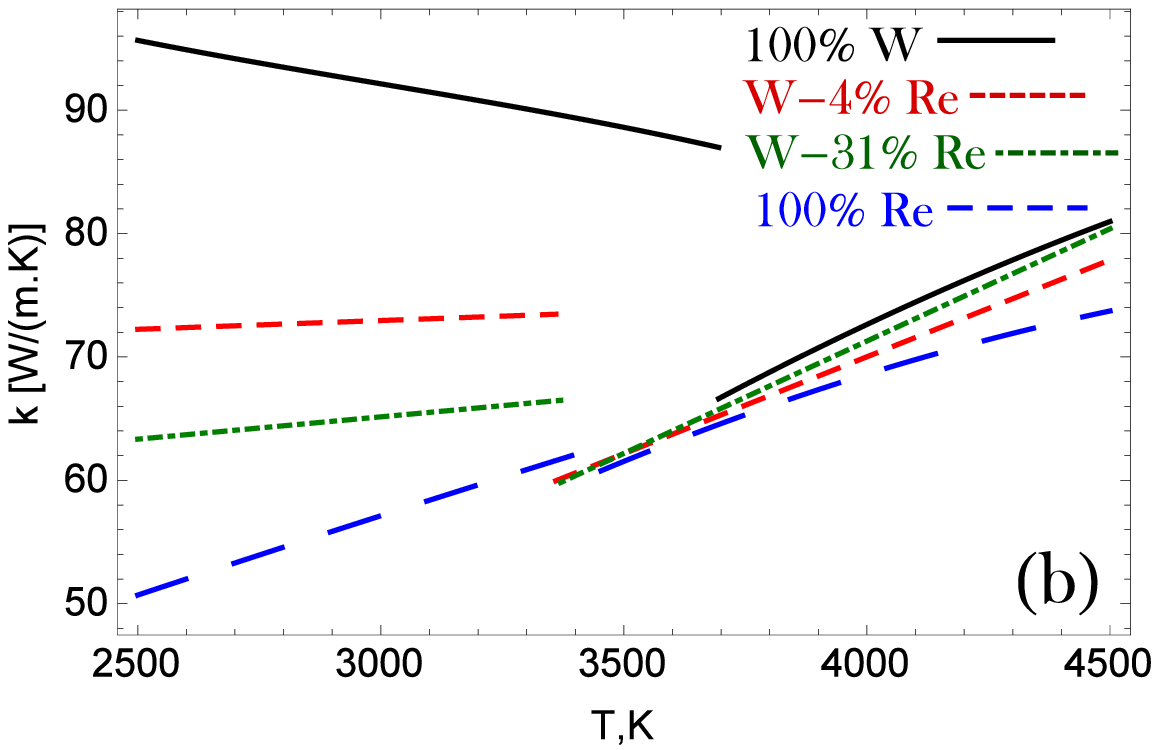}}
\caption{ (a-insert) The room temperature electrical resistivity as a function of the rhenium content for typical W-Re alloys. The measurements are adopted from Refs.\cite{handbooks04,outrorhen09} and the solid curve is drawn to guide the eye. (a-main) The thermal conductivity as a function of the temperature in the $1200-2500\,$K range for pure W and Re as well as several W-Re alloys. Pure rhenium: the recommended dataset of Ho, Powell and Liley\,\cite{conductsol2} in the range $1200-2600\,$K has been employed for quadratic polynomial fits. Rhenium alloys: the measurements of Vertogradskii and Chekhovskoi\,\cite{outrorhen05} in the range $1200-3000\,$K have been extracted from plots and fitted to quadratic polynomials. Pure tungsten: the modified Hust--Lankford fit has been employed. (b)  The thermal conductivity as a function of the temperature in the $2500-4500\,$K range for pure W and Re as well as several W-Re alloys. Pure rhenium: The tabulated experimental data of Th\'evenin, Arl\'es, Boivineau and Vermeulen\,\cite{outrorhen04} for the uncorrected electrical resistivity and the thermal volume expansion have been employed for the determination of the electrical resistivity which was then converted to the thermal conductivity with the aid of the Wiedemann-Franz law. The resulting dataset has been fitted with quadratic polynomials in the temperature ranges $2500-3453\,$K (solid) and $3453-4500\,$K (liquid). Rhenium alloys: Linear fits to the thermal conductivity measurements of Seifter, Didoukh and Pottlacher\,\cite{outrorhen08} in the temperature range $2500- 4400\,$K were employed. The alloy melting ranges are $3325-3395\,$K for W$-4\%$\,Re and $3319-3421\,$K for W$-31\%$\,Re. Pure tungsten: the recommended analytical description (the modified Hust--Lankford fit for the solid state and the Seydel--Fucke fit for the liquid state) has been employed.}\label{figureWneutronirradiation}
\end{figure*}

(d) \emph{Neutron irradiation effects.} The penetration depth of neutrons in condensed matter is several orders of magnitude larger than the penetration depth of electrons, ions or photons of comparable incident energy due to the absence of Coulomb interactions with bound electrons and the smallness of the nuclear cross-sections\,\cite{outroneutr1}. The penetration depth of fusion ions / electrons in tungsten is $\sim1-10\,$nm, whereas the penetration depth of D-T fusion neutrons in tungsten is $\sim1-10\,$cm\,\cite{outroneutr2}. Thus, neutron-induced damage is much more extended in volume than plasma-induced damage, even when accounting for bulk diffusion. Neutron irradiation can significantly modify the thermophysical properties of tungsten and particularly the thermal conductivity\,\cite{outroneutr2,outroneutr3,outroneutr4,outroneutr5,outroneutr6,outroneutr7}, as a consequence of atomic displacements (electron-defect interaction term) and nuclear transmutation (electron-phonon interaction term). The strength of this modification depends on the neutron spectrum, the neutron fluence and the irradiation temperature\,\cite{outroneutr2}. Unfortunately, experimental works on the subject are still limited\,\cite{outroneutr8,outroneutr9}. The effect of atomic displacements is hard to quantify, especially because of mitigation by annealing at high temperatures. However, a further investigation of the effect of transmutation in the tungsten power handling capabilities is viable. The principal tungsten transmutation products due to bombardment with D-T fusion generated neutrons are rhenium and osmium\,\cite{outrotransm}. At $300\,$K, the thermal conductivities of W, Re and Os are $174$, $47.9$ and $87.6$\,W/(m\,K), respectively\,\cite{conductsol2}. We shall focus on rhenium owing to its smaller thermal conductivity and its larger solubility limit in tungsten but also because tungsten-rhenium alloys have been extensively studied due to their applications in high temperature thermocouples. The solid solubility limit of Re in W increases with the temperature ($\sim28\%$ at $2000\,$K and $\sim37\%$ at $3300\,$K) and, apart from the solid solutions, two homogeneous intermetallic phases exist (broad WRe $\sigma-$phase, narrow WRe$_3$ $\chi-$phase)\,\cite{outroWRepha}. Numerous works have measured the temperature dependence of the electrical resistivity and thermal conductivity of pure Re\,\cite{outrorhen01,outrorhen02,outrorhen03,outrorhen04} as well as W-Re alloys\,\cite{outrorhen05,outrorhen06,outrorhen07,outrorhen08,outrorhen09,outrorhen10,outrorhen11} in the solid and liquid phase. Some selected datasets are illustrated in figure \ref{figureWneutronirradiation}. The generic picture emerging is summarized in the following: (i) As the Re concentration increases, the thermal conductivity of the alloy monotonically decreases. The rate of decrease is rapid up to roughly $10\%$ Re but it saturates around $20\%$. (ii) In contrast to pure  W, the thermal conductivity of solid W-Re alloys is monotonically increasing at elevated temperatures. Consequently, the thermal conductivity deviation from pure W is strongly reduced compared to the room temperature. (iii) In the liquid phase, the thermal conductivity of W-Re alloys is very close to W. In fact, even the differences between pure W and pure Re are very small above the melting points. Overall, W transmutation to Re alone can lead to a drastic reduction of the room temperature thermal conductivity up to $\sim70\%$ which becomes progressively lower as the temperature increases and eventually vanishes at the W melting point. It is worth pointing out that transmutation is estimated to be very limited in ITER but is a primary concern for DEMO\,\cite{outroneutr0}.

To sum up, the thermophysical properties of tungsten are barely affected even by strong fusion-relevant magnetic fields. Plasma contaminants and beryllium-tungsten alloying can substantially alter the W thermophysical properties but only in a relatively thin \enquote{unstable} surface layer, which implies that they can be neglected in the modelling of bulk PFCs. It should be pointed out though that the degradation of the thermal properties of such thin layers needs to be considered in the analysis of IR camera measurements\,\cite{outroplas01,outroplas02,outroplas03,outroplas04}, which might otherwise strongly overestimate the incident plasma heat flux\,\cite{outroplas01}. On the other hand, neutron irradiation can substantially modify the thermophysical properties of W in an extended volume but becomes important for high neutron fluences relevant for DEMO but not for ITER.

\subsection{Status of the experimental datasets}\label{status}

\noindent As a consequence of its extensive use in high temperature technological applications as well as due to its high melting point and very extended liquid range, pure polycrystalline tungsten can be considered as a standard reference material in the metrology of thermophysical quantities. The development of dynamic pulse calorimetry (starting from the 70s) has allowed for accurate measurements of the latent heat of fusion, the electrical resistivity, the specific isobaric heat capacity, the thermal conductivity and the mass density across the solid and liquid state. The development of levitation calorimetry (starting from the 80s) has allowed for accurate measurements of the surface tension and the dynamic viscosity at the beginning of the liquid state. Hence, it has been possible to provide accurate analytical expressions for the temperature dependence of most properties of interest based solely on experimental data and without the need for any extrapolations. The only exceptions are the surface tension and dynamic viscosity of liquid tungsten, where wide extrapolations had to be carried out beyond the melting point, since the only experimental sources on the temperature dependence referred to under-cooled liquid tungsten specimen. In spite of these limitations, the extrapolated analytical expressions performed very well against constraints imposed by rigorous statistical mechanics relations. Further measurements of the surface tension and viscosity in the unexplored temperature range are certainly desirable, but the proposed expressions are expected to be fairly accurate. Finally, the effects of plasma contamination, impurity alloying and neutron irradiation in the thermophysical properties of tungsten have been relatively poorly investigated. The sparse measurements available only account for a small part of the vast fusion-relevant parameter space, since not only these effects strongly depend on the incident plasma / impurity / neutron energies and fluences but also because they most probably operate synergetically. However, recent experiments have demonstrated that these effects can severely degrade the power handling capabilities of tungsten. Further experiments in material testing facilities are certainly required in order to evaluate the consequences for ITER and to assess the suitability of tungsten as a plasma-facing component in future fusion reactors.

\subsection{Recommended analytical expressions}\label{summary}

\noindent The thermophysical properties analyzed in this work constitute input for simulations of the thermal and hydrodynamic response of tungsten plasma-facing components, dust and droplets to incident plasma particle and heat fluxes. For this reason, in this concluding paragraph, it was judged to be more practical and convenient for the specialized reader that we simply gather the recommended analytical expressions for the temperature dependence of the thermophysical properties of pure solid and liquid tungsten. Before proceeding, it is worth mentioning that the melt layer motion code MEMOS\,\cite{introduct09,EmilspaperN} and the dust dynamics code MIGRAINe\,\cite{introduct12,outroselfi2} have already been updated following the present recommendations.

For the \textbf{\emph{latent heat of fusion}}, we recommend the typical literature value of
\begin{equation*}
\Delta{h}_{\mathrm{f}}=52.3\,,
\end{equation*}
where $\Delta{h}_{\mathrm{f}}$ is measured in kJ/mol. For the \textbf{\emph{electrical resistivity}}, we recommend the White--Minges fit in the temperature range $100\leq{T}(\mathrm{K})\leq3695$ and the Seydel--Fucke fit in the temperature range $3695\leq{T}(\mathrm{K})\leq6000$,
\begin{equation*}
\rho_{\mathrm{el}}(T)=
\begin{cases}
-0.9680+1.9274\times10^{-2}T+7.8260\times10^{-6}T^2-1.8517\times10^{-9}T^3+2.0790\times10^{-13}T^4\,\,\,\,100\leq{T}(\mathrm{K})\leq3695,\\
135-1.855\times10^{-3}(T-T_{\mathrm{m}})+4.420\times10^{-6}(T-T_{\mathrm{m}})^2\,\,\,\,3695\leq{T}(\mathrm{K})\leq6000,
\end{cases}
\end{equation*}
where $\rho_{\mathrm{el}}$ is measured in $10^{-8}\,\Omega$m or in $\mu\Omega$cm. For the \textbf{\emph{specific isobaric heat capacity}}, we recommend the White--Minges fit in the temperature range $300\leq{T}(\mathrm{K})\leq3080$, the Wilthan \emph{et al.} fit in the temperature range $3080\leq{T}(\mathrm{K})\leq3695$ and the Wilthan \emph{et al.} value in the temperature range ${T}(\mathrm{K})\geq3695$,
\begin{equation*}
c_{\mathrm{p}}(T)=
\begin{cases}
21.868372+8.068661\times10^{-3}T-3.756196\times10^{-6}T^2+1.075862\times10^{-9}T^3+\displaystyle\frac{1.406637\times10^{4}}{T^2}\,\,\,\,300\leq{T}(\mathrm{K})\leq3080\\
2.022+1.315\times10^{-2}T\,\,\,\,3080\leq{T}(\mathrm{K})\leq3695\\
51.3\,\,\,\,{T}(\mathrm{K})\geq3695
\end{cases}
\end{equation*}
where $c_{\mathrm{p}}$ is measured in J/(mol\,K). For the \textbf{\emph{thermal conductivity}}, we recommend the modified Hust--Lankford fit in the temperature range $300<T(\mathrm{K})<3695$ and the Seydel--Fucke fit in the temperature range $3695<T(\mathrm{K})<6000$,
\begin{equation*}
k(T)=
\begin{cases}
149.441-45.466\times10^{-3}T+13.193\times10^{-6}T^2-1.484\times10^{-9}T^3+\displaystyle\frac{3.866\times10^{6}}{T^2}\,\,\,\,300\leq{T}(\mathrm{K})\leq3695\\
66.6212+0.02086(T-T_{\mathrm{m}})-3.7585\times10^{-6}(T-T_{\mathrm{m}})^2\,\,\,\,3695\leq{T}(\mathrm{K})\leq6000\,,
\end{cases}
\end{equation*}
where $k$ is measured in W/(m\,K). For the \textbf{\emph{mass density}}, we recommend the White--Minges fit in the temperature range $300<T(\mathrm{K})<3695$ and the Kaschnitz--Pottlacher--Windholz fit in the temperature range $3695<T(\mathrm{K})<6000$,
\begin{equation*}
\rho_{\mathrm{m}}(T)=
\begin{cases}
19.25-2.66207\times10^{-4}(T-T_0)-3.0595\times10^{-9}(T-T_0)^2-9.5185\times10^{-12}(T-T_0)^3\,\,\,\,300\leq{T}(\mathrm{K})\leq3695\,,\\
16.267-7.679\times10^{-4}(T-T_{\mathrm{m}})-8.091\times10^{-8}(T-T_{\mathrm{m}})^2\,\,\,\,3695\leq{T}(\mathrm{K})\leq6000\,,
\end{cases}
\end{equation*}
where $\rho_{\mathrm{m}}$ is measured in g\,cm$^{-3}$ and $T_0=293.15\,$K. For the \textbf{\emph{surface tension}}, we recommend the extrapolated linear Paradis fit
\begin{equation*}
\sigma(T)=2.48-0.31\times10^{-3}(T-T_{\mathrm{m}})\,\,\,\,{T}(\mathrm{K})\geq3695\,,
\end{equation*}
where $\sigma$ is measured in N/m. For the \textbf{\emph{dynamic viscosity}}, we recommend the extrapolated Arrhenius Ishikawa fit
\begin{equation*}
\mu(T)=0.16\times10^{-3}\exp{\left(3.9713\frac{T_{\mathrm{m}}}{T}\right)}\,\,\,\,{T}(\mathrm{K})\geq3695\,,
\end{equation*}
where $\mu$ is measured in Pa\,s.

\section*{Acknowledgments}

\noindent This work has been carried out within the framework of the EUROfusion Consortium and has received funding from the Euratom research and training programme 2014-2018 under grant agreement No 633053. Work performed under EUROfusion WP MST1. The views and opinions expressed herein do not necessarily reflect those of the European Commission.

\end{document}